\documentclass[longauth]{aa} % for the long lists of affiliations 
\usepackage{graphicx}
\usepackage{txfonts}
\usepackage{lipsum}
\usepackage[export]{adjustbox}
\usepackage{arydshln}
\usepackage{xspace}
\usepackage{booktabs}

\usepackage[table, dvipsnames,svgnames,x11names]{xcolor}
 \usepackage[colorlinks=true, linktocpage, linkcolor=black, citecolor=black, urlcolor=black]{hyperref}

\newcommand{\Planck}{\textit{Planck}\xspace}

\usepackage{siunitx}
\newcommand*{\ra}[2][]{%
    \ang[
        angle-symbol-over-decimal,
        angle-symbol-degree=\textsuperscript{h},
        angle-symbol-minute=\textsuperscript{m},
        angle-symbol-second=\textsuperscript{s},
        #1]{#2}%
}

\newcommand*{\dec}[2][]{%
    \ang[minimum-integer-digits=2,angle-symbol-over-decimal,#1]{#2}%
}

\begin{document} 

   % \title{The Atacama Cosmology Telescope: An in-depth view of the $z=1.98$ galaxy cluster XLSSC~122}
   \title{XLSSC~122 caught in the act of growing up}
  \subtitle{Spatially resolved SZ observations of a $z=1.98$ galaxy cluster}

   \author{J. van Marrewijk\inst{1}
          \and
          L. Di Mascolo\inst{2,3,4}
          \and
          A. S. Gill\inst{5,6}
          \and
          N. Battaglia\inst{7}
          \and 
          E. S. Battistelli\inst{8}
          \and
          J. R. Bond\inst{9}
          \and 
          M. J. Devlin\inst{10}
          \and
          P. Doze\inst{11}
          \and 
          J. Dunkley\inst{12, 13}
          \and 
          K. Knowles\inst{14,15}
          \and
          A. Hincks\inst{5}
          \and 
          J. P. Hughes\inst{11}
          \and 
          M. Hilton\inst{16,17}
          \and
          K. Moodley\inst{17}
          \and
          T. Mroczkowski\inst{1}
          \and
          S. Naess\inst{18}
          \and
          B. Partridge\inst{19}
          \and 
          G. Popping\inst{1}
          \and
          C. Sif\'on\inst{20}
          \and 
          S. T. Staggs\inst{12}
          \and
          E. J. Wollack\inst{21}
          }

   \institute{
            European Southern Observatory (ESO), Karl-Schwarzschild-Strasse 2, Garching 85748, Germany\\ \email{joshiwa.vanmarrewijk@eso.org}.
        \and
            Astronomy Unit, Department of Physics, University of Trieste, via Tiepolo 11, Trieste 34131, Italy
        \and
            INAF - Osservatorio Astronomico di Trieste, via Tiepolo 11, Trieste 34131, Italy
        \and 
            IFPU - Institute for Fundamental Physics of the Universe, Via Beirut 2, 34014 Trieste, Italy
        \and
            David A. Dunlap Department of Astronomy \& Astrophysics, University of Toronto, 50 St. George St. Toronto ON M5S 3H4, Canada 
        \and 
            Department of Aeronautics and Astronautics, Massachusetts Institute of Technology, 77 Massachusetts Avenue, 33-207, Cambridge, MA 02139 USA
        \and
            Department of Astronomy, Cornell University, Ithaca, NY 14853, USA
        \and 
            Sapienza — University of Rome — Physics department, Piazzale Aldo Moro 5 I-00185, Rome, Italy
        \and
            Canadian Institute for Theoretical Astrophysics, University of Toronto, 60 St. George St., Toronto, ON M5S 3H8, Canada
        \and 
            Department of Physics and Astronomy, University of Pennsylvania, 209 South 33rd Street, Philadelphia, PA, USA 19104
        \and
            Department of Physics and Astronomy, Rutgers, The State University of New Jersey, Piscataway, NJ USA 08854-8019
        \and
            Joseph Henry Laboratories of Physics, Jadwin Hall, Princeton University, Princeton, NJ, USA 08544
        \and
            Department of Astrophysical Sciences, Peyton Hall, Princeton University, Princeton, NJ USA 08544
        \and 
            {Centre for Radio Astronomy Techniques and Technologies, Department of Physics and Electronics, Rhodes University, P.O. Box 94, Makhanda 6140, South Africa}
        \and
            {South African Radio Astronomy Observatory, 2 Fir Street, Observatory 7925, South Africa}
        \and
            Wits Centre for Astrophysics, School of Physics, University of the Witwatersrand, Private Bag 3, 2050, Johannesburg, South Africa
        \and
            Astrophysics Research Centre and the School of Mathematics, Statistics and Computer Science, University of KwaZulu-Natal, Durban 4001, South Africa
        \and
            Institute of Theoretical Astrophysics, University of Oslo, Norway
        \and
            Department of Physics and Astronomy, Haverford College, Haverford, PA, USA 19041
        \and
            Instituto de F\'isica, Pontificia Universidad Cat\'olica de Valpara\'iso, Casilla 4059, Valpara\'iso, Chile
        \and
            NASA/Goddard Space Flight Center, Greenbelt, MD, USA 20771
            }

   \date{Received 10 October 2023; Accepted 4 June 2024}
     \abstract
      % context heading (optional)
        {How protoclusters evolved from sparse galaxy overdensities to mature galaxy clusters is still not well understood. In this context, detecting and characterizing the hot intracluster medium (ICM) at high redshifts ($z\sim2$) is key to understanding how the continuous accretion from the filamentary large-scale structure and the mergers along it impact the first phases of cluster formation.}
      % aims heading (mandatory)
         {We study the dynamical state and morphology of the $z=1.98$ galaxy cluster XLSSC~122 with high-resolution observations ($\approx 5 \arcsec$) of the ICM through the Sunyaev-Zeldovich (SZ) effect. XLSSC~122 is the highest redshift optically confirmed galaxy cluster found in an unbiased, widefield survey.}
      % methods heading (mandatory)
        {Via Bayesian forward modeling, we mapped the ICM on scales from the virial radius down to the core of the cluster. To constrain such a broad range of spatial scales, we employed a new technique that jointly forward-models parametric descriptions of the pressure distribution to interferometric ACA and ALMA observations and multiband imaging data from ACT.}
      % results heading (mandatory)
       {We detect the SZ effect with $11\sigma$ significance in the ALMA+ACA observations and find a flattened inner pressure profile that is consistent with a noncool core classification with a significance of $\geq 3\sigma$. In contrast to the previous works, we find better agreement between the SZ effect signal and the X-ray emission as well as the cluster member distribution. Further, XLSSC~122 exhibits an excess of SZ flux in the south of the cluster where no X-ray emission is detected. By reconstructing the interferometric observations and modeling in the \textit{uv}-plane, we obtain a tentative detection of an infalling group or filamentary-like structure in the southeast that is believed to boost and heat up the ICM while the density of the gas is still low. In addition to characterizing the dynamical state of the cluster, we provide an improved SZ mass estimate $M_{500,\mathrm{c}} = 1.66^{+0.23}_{-0.20} \times 10^{14}~\rm M_\odot$.}
      % conclusions heading (optional), leave it empty if necessary 
      {Altogether, the observations indicate that we see XLSSC~122 in a dynamic phase of cluster formation while a large reservoir of gas is already thermalized.}      
    
    \maketitle
%
%-------------------------------------------------------------------

% \section{High-redshift Galaxy Clusters}
\section{Introduction}

    %generic introduction
    Despite being named for their visible galaxy constituents, the main baryonic matter component of a galaxy cluster is the thermalized, low-density plasma found between the galaxies, which is known as the intracluster medium (ICM). The traditional means of studying the ICM is through X-ray observations. In the low redshift universe ($z<1$), large samples of clusters have been observed with \textit{Chandra} \citep[e.g.,][]{2009ApJ...692.1033V} and \textit{XMM-Newton} (e.g., \citealt{Bohringer2007} \& \citealt{Arnaud2021}). At $z\gtrsim1$, however, X-ray detections become more difficult; X-ray flux falls rapidly with increasing $z$. Hence, X-ray observations of the ICM at redshifts larger than $z\gtrsim1$ usually comprise a few tens to hundreds of photons. 
    % \citep[e.g.,][]{2022A&A...667A.134T}.
    
    Fortunately, there are other means to study the ICM. When cosmic microwave background (CMB) photons propagate through the hot plasma, inverse Compton scattering shifts the blackbody spectrum of the CMB to higher frequencies. By measuring the spectral distortion, one has a direct handle on the integrated pressure along the line of sight of the hot electrons in the ICM. This phenomenon is known as the Sunyaev-Zeldovich effect (SZ effect; \citealt{Sunyaev1970}) and in contrast to X-ray observations, produces a redshift-independent surface brightness, making it ideal for high-$z$ studies of galaxy clusters. 
    
    Ground-based millimeter (mm)-wave CMB surveys have already produced large catalogs of galaxy clusters via the SZ effect containing thousands of members, such as the $>$4000 optically confirmed clusters of galaxies found in the $\sim 13,000$ deg$^2$ survey by the Atacama Cosmology Telescope (ACT) survey \citep{Hilton2021}. This catalog alone contains 222 massive ($\gtrsim 1.5 \times 10^{14}\, \rm M_\odot$) galaxy clusters at $z>1$. However, current SZ surveys, such as those with \Planck \citep{2016A&A...594A..27P}, ACT, the South Pole Telescope (\citealp[SPT;][]{2011PASP..123..568C, 2015ApJS..216...27B, 2020ApJS..247...25B}), and future CMB surveys such as those with the Simons Observatory \citep[SO;][]{2019JCAP...02..056A} and CMB-S4 \citep{cmbs4_2016, CMB-S4_2019, Raghunathan2022}, have or will have limited (arcminute-scale) angular resolutions in their main detection bands, 90\,GHz and 150\,GHz. Thus, while SZ surveys are efficient at finding clusters, they lack the spatial resolution to study the morphology and dynamical state of high-$z$ clusters, often blending a merging cluster pair into a single source \citep[e.g.,][]{DiMascolo2021}. 
    
    %Cosmology stuff
    Characterizing the dynamical state and shapes of clusters of galaxies has been pivotal for cluster cosmology \citep[e.g.,][]{Yoshida2000, Sereno2018, Ruppin2019, Lau2021}. For the past two decades, clusters of galaxies have been categorized as cool cores, morphologically disturbed (noncool cores), or simply close to the ensemble average based on their radial pressure distribution in the ICM (i.e.,\ having a pressure distribution that follows some form of self-similar or ``universal'' pressure profile; \citealt{Nagai2007}). These pressure profiles are used for measuring and interpreting the scaling relation used for inferring cluster masses. However, most SZ studies still rely on the universal pressure profile (UPP) from \cite{Arnaud2010} which is based entirely on the combination of simulations and X-ray data from the $z\lesssim0.3$ REXCESS sample. At higher redshifts, profiles are derived by stacking X-ray observations \citep{McDonald2014}. Importantly, since hydrodynamical simulations show that the pressure profile should evolve with redshift \citep[e.g.,][]{Battaglia2012, 2015MNRAS.451.3868L, Gupta2017}, it becomes increasingly inappropriate to rely on these pressure profile templates when probing higher redshifts. As upcoming telescopes such as SO and CMB-S4 are expected to be capable of detecting clusters up to redshifts of $z\sim3$ \citep{Raghunathan2022}, it becomes particularly urgent to characterize the pressure distribution of high-$z$ galaxy clusters through resolved imaging of the SZ effect.

    In practice, only a few instruments are currently capable of characterizing the ICM morphology through the SZ effect at the relevant scales (tens to a few 100 kpc) to resolve their substructure at $z>1$. Among them are MUSTANG-2 on the Green Bank Telescope \citep[e.g.,][] {Romero2020, Orlowski-Scherer2022}, the Atacama Compact Array (ACA; also known as the Morita array), the Atacama Large Millimeter/submillimeter Array (ALMA; e.g.\ \citealt{DiMascolo2021, 2023PASJ...75..311K}), MISTRAL on the SRT \citep{2023mgm..conf.1542B}, and NIKA2 on IRAM \citep[e.g.,][]{2018A&A...615A.112R,2020A&A...644A..93K, 2020A&A...642A.126R}. Observations by these instruments, at four to five times higher resolution than available with the ground-based survey telescopes used to discover SZ clusters, showed that the low-resolution signal could be resolved into separate components in higher resolution follow-up observations (for a recent review on resolved SZ studies, see \citealt{Mroczkowski2019}).

    In the most recent ACT DR5 cluster database \citep{Hilton2021}, there is only one detected cluster with a redshift $z>1.75$, namely, ACT-CL~J0217.7$-$0345 \citep[at $z = 1.98$,][]{Willis2020}, which was first discovered in the XXL X-ray survey \citep{2004JCAP...09..011P} and thus is also known as XLSSC~122 and XLSSU~J021744.1$-$034536. This work focuses on understanding the dynamical state and morphology of the ICM of this cluster, which we refer to as XLSSC~122 henceforth. Earlier measurements of the pressure distribution in the ICM from SZ studies with CARMA were presented in \citet{Mantz2014} and \citet{Mantz2018}. Very few other resolved measurements exist for clusters at $z>1.75$. These include X-ray data on four clusters that were first found in the IR/NIR: JKCS~041 \citep{Andreon2009, Newman2014}, Cl~J1449+0856 \citep{2011A&A...526A.133G}, IDCS~J1426.5+3508 \citep{Zeimann2012}, and IDCS J1433.2+3306 \citep{Stanford2012}. The first three also have follow-up SZ detections: JKCS~041 with the single-dish receiver MUSTANG-2 \citep{2023MNRAS.522.4301A}, Cl~J1449+0856 with ALMA and ACA \citep{Gobat2019}, and IDCS~J1426.5+3508 with CARMA \citep{Brodwin2012} and MUSTANG-2 \citep{2021MNRAS.505.5896A}. Beyond these, only one additional $z>1.75$ cluster has had a successful follow-up detection of its pressure profile with the SZ effect, namely the Spiderweb protocluster \citep{DiMascolo2023} via ALMA+ACA observations. We note that all five of the aforementioned clusters are outside the ACT survey footprint or below the mass limit of the ACT catalog \citep{Hilton2018, Hilton2021}.

    Yet achieving high-resolution observations is not the only aspect that needs to be considered when characterizing the pressure distribution of high-$z$ clusters of galaxies. Interferometric arrays fundamentally measure the 2D Fourier transform of the distribution of emission intensities from an astrophysical source. The long baselines provide the high-resolution samples but only short spacings can probe larger scales as the physical distance between an antenna pair is linearly proportional to the Fourier mode sampled. These Fourier modes in units of wavenumbers are thus expressed in terms of \textit{uv}-distances where $u$ and $v$ are the conventional variables for denoting the orthogonal vector basis of the Fourier space. Balancing between maximizing the collecting area for each interferometric element and mitigating concerns like the field of view size and the minimum distance between two antennas to avoid a collision imposes a fundamental constraint on the minimum length a baseline can be, which inevitably leads to an incomplete \textit{uv}-coverage and strong spatial filtering. To address the loss of information at large angular scales ($>1 \arcmin$), we can turn to data from bolometer arrays on single-dish telescopes. However, existing single-dish facilities also do not provide the extent of complete Fourier sampling required for high spatial dynamic range and unbiased image reconstruction \citep{Frayer2017,Plunkett2023}. In order to address this missing baseline problem in the ALMA and ACA data, we will also make use of known radial pressure profiles to interpolate over the missing information.
        
    In this paper, we treat this issue by combining archival SZ observations from both the main ALMA (12m-array) and ACA (7m-array) with single-dish data from ACT. This comprehensive approach enables us to probe the broad range of spatial scales needed to provide the first detailed, sub-arcminute view of the ICM in the highest redshift cluster found in the current generation of SZ survey data. The remainder of this work is outlined as follows: we provide an overview of the measurements used for our analyses and the data reduction details in Section~\ref{sec:obs}. Section~\ref{sec:methods} provides the methodology on how the forward modeling is implemented. There are three results sections: The first, Section~\ref{sec:interloper_method}, is on how we handle and correct for compact source contamination; Section~\ref{sec:results} describes the first results on the derived pressure profiles; And finally, Section~\ref{sec:twocomp} goes into detail on how to recover and model asymmetric surface brightness distributions. The implications of our observations and an exploration of their potential interpretations are provided in the following two sections. Section \ref{sec:masses} treats the derived halo mass of XLSSC~122 and Section~\ref{sec:discussion} the morphological implications of the results obtained from the forward modeling. Finally, we summarize and conclude our work in Sect. \ref{sec:conclusion}.
    
    For all calculations, the assumed cosmology is based on \citet{Planck2014_XIII},  a spatially flat, $\Lambda $ cold dark matter ($\Lambda$CDM) model with $H_0=67.7$ kms$^{-1}$Mpc$^{-1}$ and $\Omega_{\rm m} = 0.307$. Here $1\arcsec = 7.855$ kpc for XLSSC~122 ($z=1.98$).

\section{Observations of XLSSC~122}\label{sec:obs}
    
    The galaxy cluster XLSSC~122 has not gone unnoticed. It was discovered via its extended X-ray emission in the \textit{XMM-Newton} Large Scale Structure survey \citep{2004JCAP...09..011P}, and first named and described in \citet{Willis2013}, almost a decade later. XLSSC~122 was further followed up with CARMA observations that map the ICM through the SZ decrement at 30 GHz \citep{Mantz2014} as well as a combined analysis of the CARMA plus short \textit{Chandra} and deeper \textit{XMM-Newton} X-ray follow-up observations \citep{Mantz2018}. At the same time, \cite{Hilton2018} noted that while not meeting the threshold for detection in the ACTpol sample, the data showed a 3.5$\sigma$ decrement at the location of this cluster, which if verified would correspond approximately to the mass reported in \cite{Mantz2014}, namely $M_{\rm 500,c} \simeq 1\times10^{14}~M_\odot$. The cluster can be found at an RA and Dec of \ra{2;17;44.2128}, \dec{-3;45;31.68}.
    The cluster was formally detected at a signal-to-noise ratio $S/N>5$ in the next ACT cluster catalog \citep{Hilton2021}.
    
    Optical spectroscopy using the \textit{Hubble Space Telescope} (HST) Wide Field Camera 3 established that 37 galaxies are associated with XLSSC~122 \citep{Willis2020} and found the average spectroscopic redshift to be $z=1.98$. Since then, studies on the cluster member properties using multiwavelength archival data came to light, finding that $88\substack{+4\\-20}\,\%$ of the members of the cluster lying within $0.5\,r_{500,\mathrm{c}}$\footnote{$M_{500,\mathrm{c}}$ is defined as the total mass of the cluster within a radius $r_{500,\mathrm{c}}$ in which the average density is 500 times the critical density of the Universe at that redshift.} are quenched and exhibit a larger half-light radius than field galaxies \citep{Noordeh2021}. The presence of quenched galaxies, and thus the existence of a red sequence, plus the detection of the ICM indicates that XLSSC~122 is, in some sense, already a mature cluster.
    
    To this multiwavelength view of XLSSC~122, we add microwave data from ALMA (the main array of 12m antennae), ACA (the 7m-array), and ACT Data Release 6 (DR6; \citealt{2023arXiv230701258C}) to map the SZ decrement from the core to the virial radius at sub-arcminute resolution. These observations will map any asymmetry in the pressure distribution of the ICM and characterize its radial profile; both were previously inaccessible because of the low resolution ($\approx 1\arcmin$) and low $S/N$ of the previous data. In the following subsections, we will describe ALMA, ACA, and ACT observations that will determine if this cluster is thus truly relaxed and mature or is still actively forming.

    \subsection{The Atacama Large Millimeter/Submillimeter Array}

        We rely upon archival yet unpublished ALMA and ACA Band 3 observations (henceforth denoted ALMA+ACA) made on 2016-10-22 and 2017-01-06, respectively, for characterizing the SZ decrement. For both observations with project code 2016.1.00698.S (PI: A. Mantz), calibrated datasets were obtained through the calibrated measurement set service \citep[CalMS;][]{Petry2020} of the European ALMA Regional Center using CASA 5.4.0. The total on-source time for the ACA and ALMA observations are 3.9 and 0.42 hours, respectively.
        
        \begin{figure}[t]
            \centering
            \includegraphics[width=0.98\hsize]{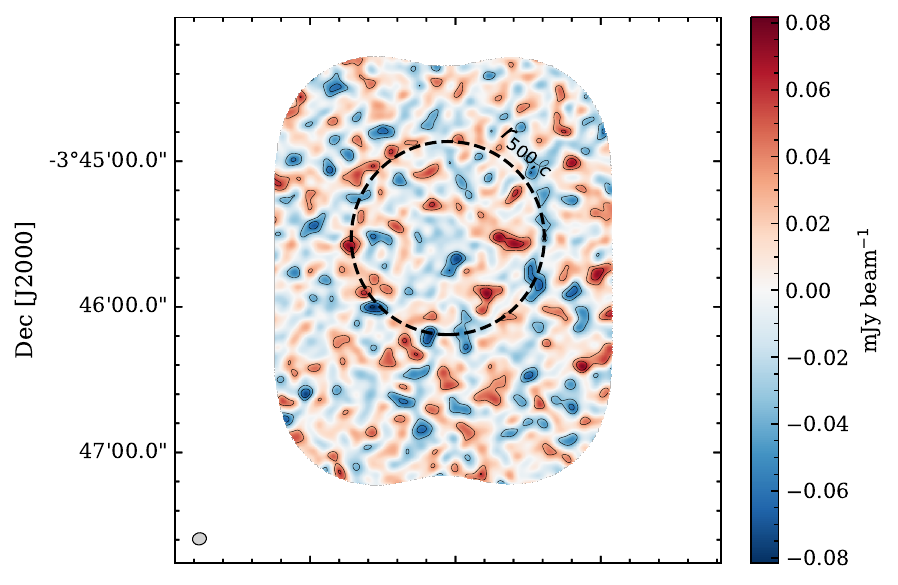}
            \includegraphics[width=0.98\hsize]{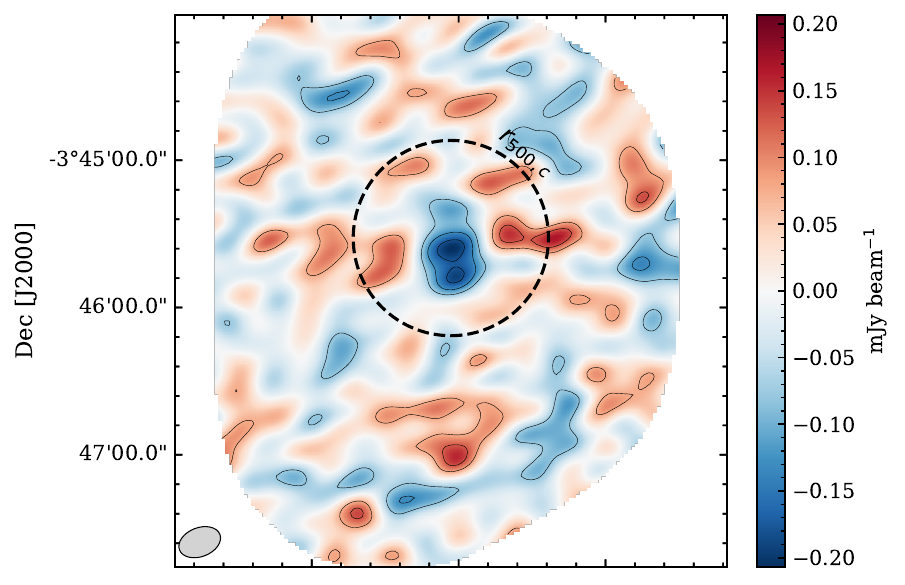}
            \includegraphics[width=0.98\hsize]{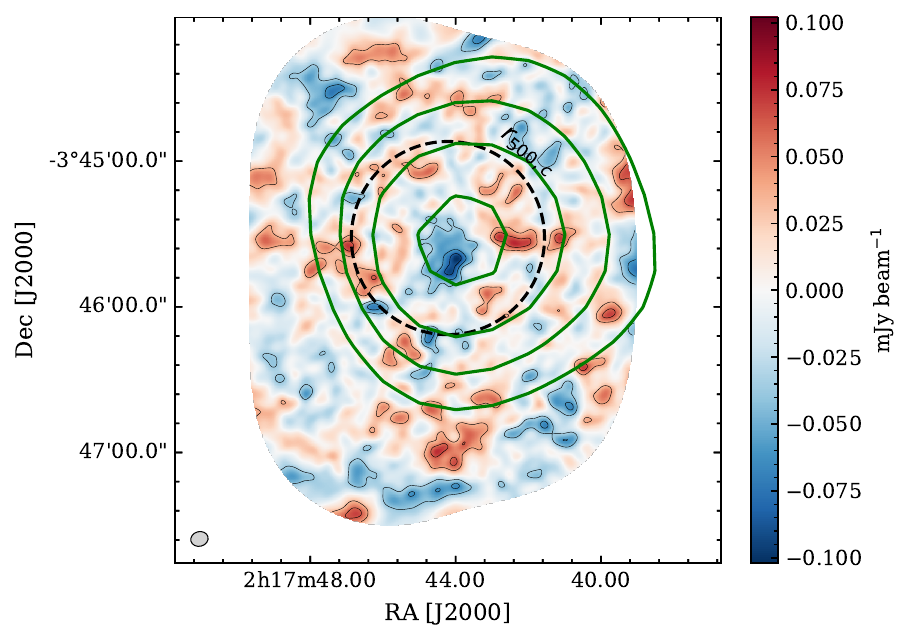}
            \vspace{-2mm}
            \caption{ALMA observations of XLSSC~122. From top to bottom, we show the raw (dirty) continuum images of the 12m-array mosaic, ACA mosaic, and the jointly imaged Band 3 observations. The latter image (bottom panel) is tapered with a \textit{uv}-taper of 20~$\rm k\lambda$. Black contours in each panel are drawn at [$-4.5$, $-3.5$, $-2.5$, $-1.5$, 1.5, 2.5, 3.5]$-\sigma$ except for the top panel, which excluded the $\pm~1.5\sigma$ contours. We find a central noise RMS of $\sigma=0.051$, $0.014$, $0.022$ mJy~beam$^{-1}$, respectively. The dashed circles in all panels indicate $\rm r_{\rm 500,c}$ centered on the BCG. The green contours in the bottom panel indicate the ACT-$y$ map contours and are drawn at [2.5, 3.5, 4.5, 5.5] times the local noise level. The beams of the ACA, ALMA, and jointly imaged ACA+ALMA observations are indicated in the lower-left corner of each panel. We clearly see that the ACT and ALMA+ACA observations align spatially. }
            \label{fig:combined}
            \vspace{-2mm}
        \end{figure}

        \begin{figure}[t]
            \centering
            \includegraphics[width=\hsize]{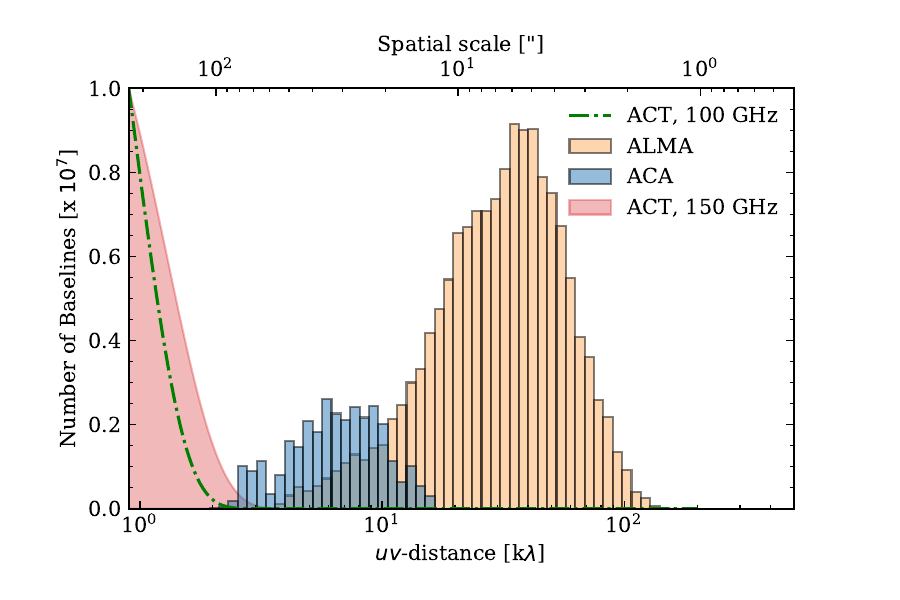}
            \caption{Distribution of the number of visibilities as a function of the \textit{uv}-distance sampled with ALMA (yellow) and ACA (blue) Band-3 observations. In red, we overlay the ACT beam at 150 GHz ($\sim 1\farcm4$) and 100 GHz ($\sim 2\farcm0$) as a green dashed line. The y-axis is scaled using arbitrary values regarding the ACT beams.}
            \label{fig:uvcoverage}
        \end{figure}
                
        \begin{figure}[t]
            \centering
            \includegraphics[width=0.87\hsize]{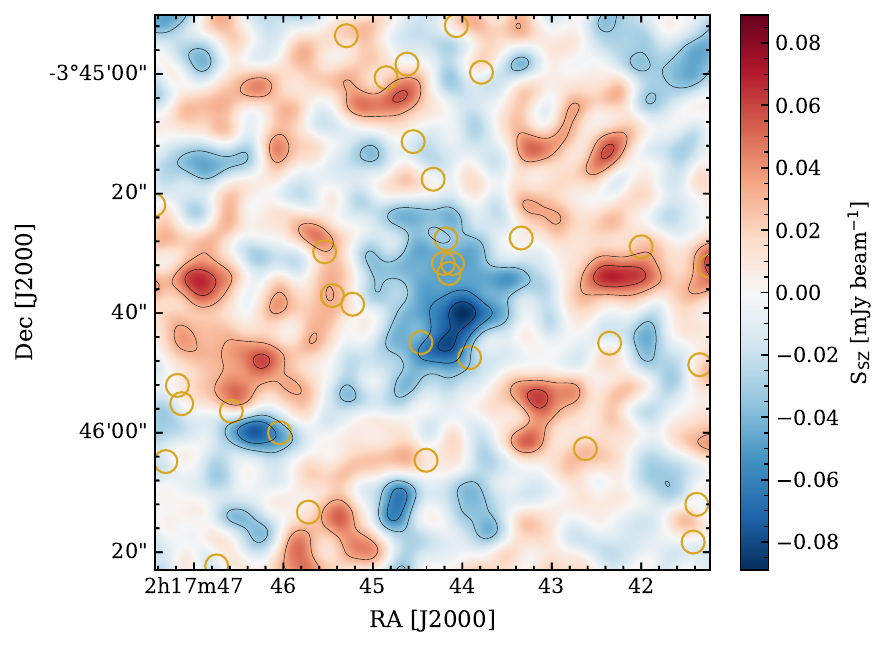}
            \includegraphics[width=0.87\hsize]{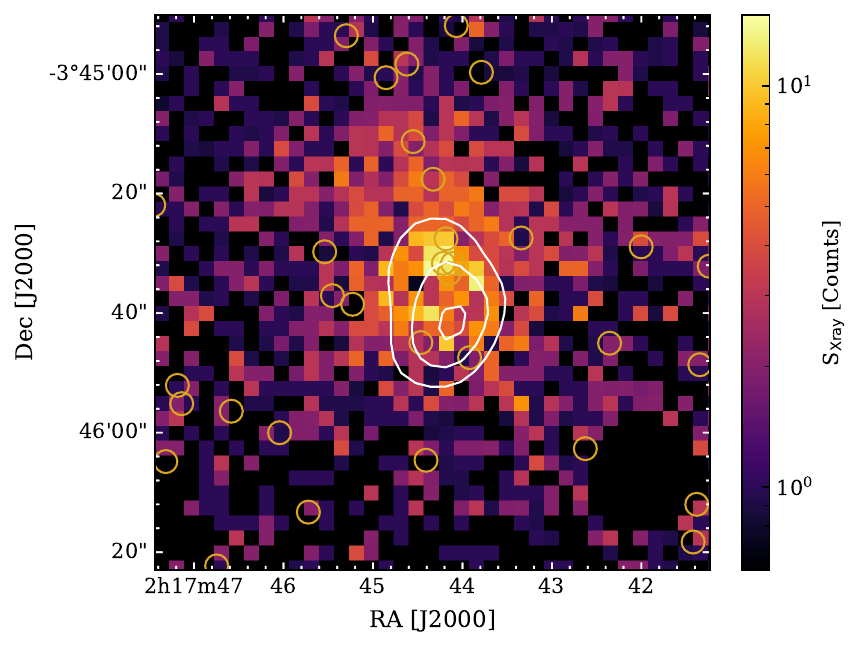}
            \includegraphics[width=0.87\hsize]{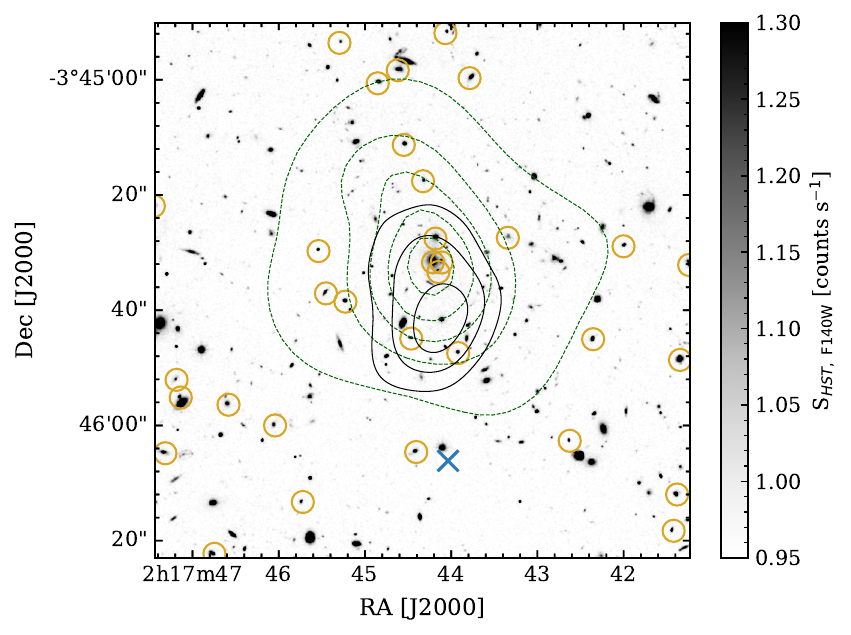}
            \vspace{-2mm}
            \caption{Multiwavelegth view of XLSSC~122. \textit{Top:} A dirty map of the joint ALMA+ACA Band 3 observations overlaid with the distribution of member galaxies of the cluster shown as small gold circles, as in all other panels. The black contours are drawn at [-4.5, -3.5, -2.5, -1.5, 1.5, 2.5, 3.5]$-\sigma$. \textit{Middle:} \textit{XMM-Newton} image of \citet{Mantz2018} overlaid with the directly visible SZ decrement as seen by the ALMA+ACA imaging with a \textit{uv}-taper of 10~k$\lambda$. \textit{Bottom:} HST imaging from \citet{Willis2020} used to find the cluster members. Here, we overlay in black the directly visible SZ decrement as seen by the ALMA+ACA observations, and in dark green, we show the adaptively smoothed X-ray contours drawn at [2, 4, 6, 8, 10]$-\sigma$. We further highlight the SZ centroid derived from CARMA measurements \citep{Mantz2018} with the blue cross. Overall, this figure indicates a coherence between the pressure and density distribution of the ICM seen by ALMA+ACA and X-ray measurements, and the cluster member distribution.} \vspace{-10mm}
            \label{fig:multiwavelength}
        \end{figure}

        Both the 12m-array and 7m-array observations span four spectral windows: 89.51--91.50, 91.45--93.44, 101.51--103.49, and 103.51--105.49\,GHz, observed in continuum mode. The spectral windows are set up to exclude any strong molecular emission lines of cluster members (we note, however, that we identified spurious emission lines of other galaxies along the line of sight; see Sec.\ \ref{sec:interloper_method} below for details). Both observations were carried out in mosaic mode, with the ACA observations consisting of five pointings and the ALMA observations consisting of 11. The ACA and ALMA observations reach a central continuum sensitivity of $0.24$ and $0.037$\,mJy~beam$^{-1}$ per pointing with an angular resolution of $11\arcsec$ and $2\arcsec$, respectively. The maximum recoverable scales of the ACA and ALMA observations are $77\arcsec$ and $23\arcsec$ per pointing. The ACA and ALMA continuum maps, as well as a jointly imaged ALMA+ACA map, are shown in Figure~\ref{fig:combined}. The maps show dirty images in which the true sky is convolved with the transfer function, which arises from the incomplete uv-coverage of the interferometer. The bottom panel and remaining figures in this paper that show ALMA+ACA observations are tapered\footnote{Tapering is equivalent to smoothing the PSF with a Gaussian. However, tapering in \textit{uv}-space uses a natural weighting scheme that down weights higher spatial frequencies relative to lower spatial frequencies to suppress artifacts arising from poorly sampled areas of the uv-plane.} to have a beam with an FWHM of $5\arcsec$.
        
        Furthermore, we make use of ancillary ALMA Band 4 observations (PI: J. Zavala, 2018.1.00478.S) to get better spectral constraints on the contaminating emission from dusty galaxies. These observations, with project code 2018.1.00478.S, were performed on 2019-01-18 and reduced via the standard CASA 5.4.0 pipeline. The Band 4 observations span four spectral windows -- 139.7--141.6, 143.0--143.9, 153.1--154.0, and 154.9--155.9\,GHz -- with a single pointing centered on $\mathrm{RA} =\ra{02;17;42.8}$, $\mathrm{Dec}=\dec{-03;45;31.2}$. It reaches a continuum sensitivity of $\sim0.024$ mJy~beam$^{-1}$ RMS.
    
    \subsection{The Atacama Cosmology Telescope}\label{ssec:act}
    
        To characterize larger angular scales of XLSSC~122, and in particular, to constrain the integrated Compton $Y$ ($=\int y d\Omega$) (see Sec.~\ref{sec:sz_def}), we include data from the Atacama Cosmology Telescope (ACT). ACT was a 6-meter, off-axis Gregorian survey telescope that operated from 2008 to 2022 in the Atacama Desert of Chile. Over its lifetime, the detectors and receiver were modified such that it covered, at various times, frequency bands centered at approximately 30, 40, 100, 150, 220, and 280\,GHz; detectors were sensitive to polarization starting in 2013 \citep{Swetz_2011,Thornton_2016,henderson2016advanced}. The receiver consisted of three arrays of detectors, each in its own optics tube, such that each was located at a different position on the focal plane. For most of the period of observation used in this paper, each array was sensitive to two different frequency bands: thus, at any given time, six array--frequency pairs were observed.

        In this paper, we use the ACT maps at 100 and 150\,GHz from DR6 with a point-source sensitivity integrated over the beam of 0.5\,mJy and 0.8\,mJy, respectively. These observations are taken from 2017 to 2022. We note that these are the frequencies that are most sensitive to the SZ effect (see Sec.~\ref{sec:sz_def}). ACT provides the larger scales with a FWHM resolution of $\approx2'$ and $\approx1.4'$ at 100 and 150\,GHz, respectively. The bottom panel in Figure \ref{fig:combined} overlays the contours of the ACT DR6 $S/N$ map of XLSSC~122 to be included in Hilton et al.\ (in prep.). The figure shows that XLSSC~122 is largely unresolved by the ACT observations. 

        Figure \ref{fig:uvcoverage} overlays the baseline distributions for the spatial scales sampled by ALMA and ACA with the ACT beams at 100 and 150\,GHz to show which spatial scales we are sensitive to when utilizing both types of observations. Here, angular scales are converted to \textit{uv}-distances through the relation:
        
        \begin{equation}
            uv\text{-distance} [\lambda] =\frac{1}{\text{angular scale}~[\text{radians}]}.
        \end{equation}

        \noindent Thus, the scale probed in radians equals the inverse of the baseline length in wavenumbers.
        
    \subsection{Auxiliary data comparison}

        Throughout this work, we will compare the results obtained from our analysis of the SZ effect with the X-ray surface brightness distribution. As mentioned above, XLSSC~122 was first discovered via its extended X-ray emission as seen with \textit{XMM-Newton}. In total, \textit{XMM-Newton} collected $\approx 1096$ source photons in an exposure of 106 ksec, while higher-resolution \textit{Chandra} observations only collected $\approx 200$ source counts in 182.2 ksec after light curve filtering (deflaring). Here, we use the previously unpublished data from the archive (PI: Noordeh)\footnote{{\it Chandra} Observation IDs 22562, 22563, 22857, 22868, 22869, and 22870.} and applied the standard ACIS reprocessing technique in order to merge the \textit{Chandra} observations into a flux-corrected image using the \texttt{ciao} tools. However, the combined \textit{Chandra} data translates to $\lesssim 4$ source photons per hour of observation time, which is not uncommon in high-redshift observations of clusters of galaxies in the X-ray regime. As an example, X-ray observations of the Spiderweb protocluster collected approximately 1 photon per 3.6 ksec \citep{2022A&A...667A.134T}.
        % It should be noted that future observations will present even greater challenges as \textit{Chandra} is experiencing a rapid loss of sensitivity at low photon energies, which is where the majority of the signal will be redshifted \citep{2018SPIE10699E..6BP}. 
        Therefore, to study the diffuse ICM in XLSSC~122 from the X-ray perspective, we henceforth only rely on the \textit{XMM-Newton} observations.
         
        Figure \ref{fig:multiwavelength} compares the SZ decrement (also shown in Figure \ref{fig:combined}) with the observed X-ray emission seen by \textit{XMM-Newton} from \citet{Mantz2014, Mantz2018} and the deep optical HST-imaging from \citet{Willis2020}. Our mm-wave observations are in contrast with the findings of \citet{Mantz2018}, who reported an offset of $35\arcsec \pm~ 8\arcsec$ ($295 \pm 64$ kpc at $z=1.98$) between the X-ray emission and the CARMA-measured SZ decrement (its centroid is marked with the blue cross in the bottom panel of Fig.~\ref{fig:multiwavelength}). However, the CARMA measurements suffer from low signal-to-noise ($\sim5\sigma$) and low resolution, which makes it difficult to account for and remove contaminating sources.
        The ALMA+ACA observations, on the other hand, clearly reveal the central SZ decrement, while the higher resolution better allows us to mitigate the impact off-centered point sources have on the overall SZ measurement. 
        
\section{Methodology}\label{sec:methods}

    To study the dynamical state and morphology of XLSSC~122, we forward-model the pressure distribution of the ICM to the observations. In this section, we describe how we jointly model interferometric ACA and ALMA measurements with multi-band imaging data from ACT. In brief, the routine flows as follows: 
    (1) From a parametric description of the pressure distribution, we map the ICM on a three-dimensional grid which is then integrated along the line of sight to produce a projected Compton $y$ map (see section \ref{sec:parametric}).
    (2) Using the spectral scaling of SZ effect, we convert the Compton $y$ model to a surface brightness distribution for each of the spectral bands of the analyzed data (see section \ref{sec:sz_def}).
    (3) We Fourier transform the surface brightness map to the visibility space, which is the native space of the ALMA and ACA data. We then apply the ALMA and ACA transfer function to the model to account for the missing baselines and compute the likelihood of the resulting model on the unbinned visibilities assuming Gaussian properties for noise statistics (see section \ref{sec:visibillity_Modeling}).
    (4) As the ACT and ACA+ALMA observations probe different spatial scales, we can forward model the same pressure distribution to both observations separately and combine them by adding the log-likelihoods linearly (see section \ref{sec:act_method}). We use the radial dependency of the parameterized model to fill in the gap in the angular scales where the observations are less sensitive. 

     \subsection{Parametric descriptions of the ICM}\label{sec:parametric}
        \subsubsection{Pressure profiles}

            Our forward modeling technique Fourier transforms the SZ signal based on a parametric description of the ICM to the \textit{uv}-plane. As a parameterized model, we choose the generalized Navarro-Frenk-White (gNFW) profile proposed by \citet{Nagai2007}:
            
            \begin{equation}\label{eq:gNFW}
                P_{\rm e} (r) = P_{\rm e,i} \left(\frac{r}{r_{\rm s}}\right)^{-\gamma} \left[ 1+ \left(\frac{r}{r_{\rm s}}\right)^{\alpha} \right]^{(\gamma-\beta)/\alpha},
            \end{equation}
            
            \noindent where $P_{\rm e,i}$ is the pressure normalization and $\gamma$, $\alpha$, and $\beta$ the shape parameters of the broken power law. Respectively, they represent the slopes at small, intermediate, and large radii with respect to the characteristic radius $r_{\rm s}$. In this description, the pressure distribution is a function of the radial distance $r$ from the centroid of the cluster ($x_{\rm gNFW}$, $y_{\rm gNFW}$). Eq. \eqref{eq:gNFW} is often rewritten to break the spherical symmetry and encapsulate a projected eccentricity which we define as

            \begin{equation}
                e = 1-\frac{b}{a},
            \end{equation}

            \noindent with $a$ and $b$, the major (with $a=r_{\rm s}$) and minor axis of the ellipse, respectively.

            Further, the theoretical formalism of \citet{Nagai2007} can be rewritten such that the normalization parameter $P_{\rm e,i}$ is linked to the halo mass of the cluster, $M_{500,\mathrm{c}}$, through the self-similarity principle, the local $M_{500,\mathrm{c}}-Y_{\rm X}$ relationship based on the REXCESS sample, and the redefinition of $r_{\mathrm{s}}$ such that $(r/r_\mathrm{s})$ is expressed in terms of the concentration parameter $c_{500,\mathrm{c}}$, such that
            
            \begin{equation}\label{eq:concentration}
                \left(\frac{r}{r_{\rm s}}\right) = c_{500,\mathrm{c}} x,
            \end{equation}
            
            \noindent with $x = r/r_{500,\mathrm{c}}$ and $c_{500,\mathrm{c}} = r_{500,\mathrm{c}}/r_{\rm s}$. This way, the amplitude of the pressure profile and also its specific radius is scaled according to an additional self-similar principle and is a function of halo mass ($M_{500,\mathrm{c}}$) and redshift ($z$) via the $c_{500,\mathrm{c}}-M_{500,\mathrm{c}}$ relationship. This formalism was introduced by \citet{Arnaud2010} and provides additional constraints in the parameter space, simplifying the modeling of the gNFW. All combined, this empirical relation is expressed as

            \begin{equation}\label{eq:A10_form}
                P_{\rm self-similar}(M_{\rm 500,c},z) = P_e(r) \times P_{500,c}, 
            \end{equation}

            \noindent with $P_e(r)$ as described in Eq.~\eqref{eq:gNFW} but with the substitution provided in Eq.~\eqref{eq:concentration}. Following \citet{Arnaud2010}, the mass dependency comes from the $P_{500,c}$ parameter which is defined as

            \begin{equation}
                P_{500,c}(M_{500,c}, z) = 1.65 \times 10^{-3}~E(z)^{8/3} \left[\frac{M_{500,c}}{3\times10^4 M_\odot}\right]^{2/3+a_p(r/r_s)}
            \end{equation}

            \noindent in units of keV cm$^{-3}$. Here, $E(z)$ is the ratio of the Hubble constant at redshift $z$ to its present value $H_0$ and $a_p(r)$ the parameter that accounts for deviations from self-similarity in the core of galaxy clusters:

            \begin{equation}
                a_p(r/r_s) = a_0/[1 + 8\times(c_{500,c}x)^3].
            \end{equation}
            
            Since \citealt{Arnaud2010} (hereafter, A10), the A10 profiles have been used throughout the literature as the universal pressure profiles \citep[e.g.,][]{Arnaud2010, McDonald2014} by estimating averaged values of $\gamma$, $\alpha$, $\beta$, and $c_{500,\mathrm{c}}$ for different dynamical states, halo masses, and redshifts.
            
            In this work, we will construct models with both the gNFW-formalism as presented in Equation~\eqref{eq:gNFW} and the parameterization described by \citet{Arnaud2010}. 
            When we use the formalism of \citet{Arnaud2010}, we will freeze the shape parameters ($\alpha$, $\beta$, $\gamma$, and $c_{500}$) of the derived classifications at the averaged values derived from \citet{Arnaud2010} and \citet{McDonald2014}. These values are split in subsamples referring to local ($z<0.2$) and more distant ($0.6<z<1.2$) clusters, respectively. In the remainder of this work, we refer to each of these two formalisms as the theoretical formalism (referring to Eq.~\ref{eq:gNFW}) and the empirical formalism (referring to Eq.~\ref{eq:A10_form}), respectively. 
            
            Both \citet{Arnaud2010}~\&~\citet{McDonald2014} split their cluster samples into three classifications: the cool cores (CC), the morphologically disturbed (MD, also known as noncool cores), and the ensemble-averaged classification (also known as the universal pressure profiles, abbreviated to UPP). Cool cores are, on average, more relaxed clusters of galaxies with a cusped core, while morphologically disturbed profiles are a direct consequence of merger activity and exhibit a more flattened inner pressure profile. Hence, by modeling pressure profiles that correspond to these different classifications for the two redshift bins, we can classify the dynamical state of XLSSC~122.
           
        \subsubsection{The thermal Sunyaev-Zeldovich effect}\label{sec:sz_def}
        
            To model the hydrostatic properties of the ICM and link it to observables in the mm-wave regime, we use the SZ effect \citep{1969Ap&SS...4..301Z, Sunyaev1970, 1972CoASP...4..173S}. When CMB photons propagate through the hot plasma in the ICM, inverse Compton scattering shifts the photons to higher frequencies, distorting the blackbody (BB) spectrum of the CMB. The predominant source of the transformation, called the thermal SZ effect, is caused by the pressure distribution of the hot electrons in the ICM. This frequency-dependent distortion is the observable we exploit in this work and which we henceforth refer to simply as the SZ effect. (For a thorough description of the various types of other SZ effects, we refer the reader to the review paper of \citealt{Mroczkowski2019}.)
            
            The amplitude of the SZ effect is a function of the Compton-$y$ parameter:
            
            \begin{equation}
                y \equiv \int \frac{k_\textsc{b} T_\mathrm{e}}{m_\mathrm{e} c^2}~ \text{d}\tau_\mathrm{e} = \int \frac{k_{\rm B} T_\mathrm{e}}{m_\mathrm{e} c^2} n_\mathrm{e} \sigma_\textsc{t}~ \text{d}l = \frac{\sigma_\textsc{t}}{m_\mathrm{e}c^2} \int P_\mathrm{e}~ \text{d}l~ .
            \end{equation}
    
            \noindent Here, $\sigma_\textsc{t}$ is the Thomson cross section, $P_\mathrm{e} =n_\mathrm{e} k_\textsc{b} T_\mathrm{e}$ is the thermal pressure due to the electrons, $n_\mathrm{e}$ is the number density of the electrons, $k_\textsc{b}$ is Boltzmann's constant, $T_\mathrm{e}$ the electron temperature, $m_\mathrm{e}$ the electron mass, $c$ the speed of light, $\tau_\mathrm{e}$ the opacity, and $l$ the path along the line of sight. Thus the magnitude of the SZ signal is a direct measure of the integrated line of sight pressure and can be obtained by integrating along $\mathrm{d}l$ with the radial dependence of $P$ given by Eq. \eqref{eq:gNFW}. Finally, the corresponding distortion signal is given by
            
            \begin{equation}\label{eq:spectrum}
                \Delta I_{\nu} \approx I_0 y \frac{x^4 \text{e}^x}{\left(\text{e}^x-1\right)^2}\left(x\frac{\text{e}^x+1}{\text{e}^x-1}-4\right) \equiv I_0 y g(x),
            \end{equation}
            
            \noindent in terms of the CMB intensity $I_0 \approx 270.33 ~ \left(T_{\textsc{cmb}}/2.7255~ \text{K}\right)^3$ MJy~sr$^{-1}$ and $x = h\nu~(k_\textsc{b} T_{\textsc{cmb}})^{-1}  \approx \nu / 56.8~\rm GHz$, where the CMB temperature is adopted from \citet{2009ApJ...707..916F}.

            As shown in Eq. \eqref{eq:spectrum}, the SZ distortions are frequency dependent. In ACT and ALMA's main SZ-detection bands ($\sim 90$ and $150$ GHz) the distortion spectrum becomes negative, manifesting as a negative surface brightness in the continuum maps. The equations above assume nonrelativistic speeds for the electrons. We note, for instance, that in a galaxy cluster with a mean ICM temperature of 2 keV, the relativistic correction term would change the overall amplitude of the SZ effect at 100 GHz by $1.3\%$, and at 4 keV by  $2.6\%$. At 150 GHz, the 4 keV correction is 2.8$\%$. Therefore, any relativistic corrections fall within the calibration uncertainty of the ALMA and ACA observations ($\sim5\%$). As we do not expect the electron temperature in XLSSC~122 to exceed 5 keV \citep[see][]{Mantz2018, Duffy2022}, we have neglected the temperature-dependent relativistic corrections of the thermal SZ in this work.

    \subsection{Visibility-based modeling of the ICM}\label{sec:visibillity_Modeling}

    \subsubsection{The visibility plane}\label{sec:visibillity_Modeling}

        We employ a forward-modeling approach to determine the best-fitting model parameters that describe the ALMA interferometric and ACT map-domain data on XLSSC~122. This involves reconstructing the surface brightness distribution by forward-modeling the pressure distribution of the hot electrons in the ICM to both types of observations. Here, we rely on a \textit{uv}-space Bayesian approach based on the work of \citet{DiMascolo2019a}. For details, we refer to the original presentation, discussion, and references therein; here, we provide a brief summary. 
        
        The reconstruction method makes use of the native measurement space of interferometric data, which incompletely samples the \textit{uv} (i.e., Fourier)-plane. Interferometric observations measure Fourier transforms of the distribution of emission intensities from an astrophysical source at a given angular scale and spectral resolution. The Fourier-space measurements, called visibilities, have coordinates $\left(u,v\right)$ representing the projected baseline distances between two antennas in a plane normal to the direction of the phase reference position \citep[see e.g.,][page 91]{thompson1986interferometry}. Therefore, each visibility $\mathcal{V} \left(u,v\right)$ is defined as

        \begin{equation}\label{eq:visibilities}
            \mathcal{V} \left(u,v\right) = \int_{-\infty}^{\infty} \int_{-\infty}^{\infty} \frac{A_N\left(l,m\right) I\left(l,m\right)}{\sqrt{1-l^2-m^2}} e^{-2\pi i\left(ul+vm \right)} {\rm d}l \ {\rm d}m,
        \end{equation}
        
        \noindent with $A_N\left(l,m\right)$ being the normalized primary beam pattern of the antennas (assumed to be the same for each), and $I\left(l,m\right)$ the source intensity distribution.
        Infinitesimal terms d$l$ and d$m$ in this equation combine to form a solid angle, ${\rm d}\Omega = {\rm d}l \ {\rm d}m / \sqrt{1- l^2 - m^2}$, such that the power received by each antenna is $P = \int A_N(l,m)I(l,m) \Delta\nu {\rm d}\Omega$, where $\Delta\nu$ is the instrumental bandpass (and we make the approximation that $A_N$ is the same across the bandpass).
        
        Modeling in the \textit{uv}-plane avoids issues related to the deconvolution of interferometric data (e.g.\ heavily correlated image-space noise, the filtering of large spatial scales, and the nonuniform weighting of the signal across the baselines); it also takes full advantage of the knowledge of the exact visibility sampling function as we interpolate the Fourier transformed ICM model to match the \textit{uv}-coordinates of the ACA+ALMA observations. Regarding handling the ALMA+ACA mosaic, we model each pointing individually and correct for the primary beam attenuation per field. The fields are combined at the likelihood level of our routine, adopting a Gaussian likelihood. The implementation uses the static nested sampling method implemented in the \texttt{dynesty} package \citep{Speagle2020Dynesty, sergey_koposov_2023_7600689}.

    \subsubsection{Joint likelihood modeling of the ICM via ALMA, ACA, and ACT observations}\label{sec:act_method}

        The remainder of this section will describe how we utilize both types of observations to model the SZ effect over a broad range of spatial scales, how we correct for the different covariances between the various data sets, and how we infer the significance of the modeling via the Bayes factor.

        Similar to how we combine the separate pointings in the ALMA and ACA mosaics, we also combine the ALMA+ACA interferometric observations with the ACT maps at the likelihood level. Thus, in every iteration, we model the pressure distribution given a set of parameters, project the model to the \textit{uv}-plane, apply the transfer function for each specific set of observations, and compute the Gaussian log-likelihoods of the ACA+ALMA and the ACT observations individually. We treat the ACT and ACA+ALMA observations as independent when computing the final posterior distribution of the model parameters. This is done by adding the log-likelihoods linearly. Hence, we assume no covariance between the ACA+ALMA observations and the ACT maps. This approximation is justified by the minimal overlap in angular scales probed by the two datasets, shown in Figure~\ref{fig:uvcoverage}.
        
        Even though the covariance between ACT and ALMA+ACA observations can generally be neglected, there can be covariance between the ACT maps from different frequency--array pairs (see Sec.~\ref{ssec:act}), particularly since the primary CMB anisotropies---which are a source of noise for us---are present in them all.\footnote{We note that any CMB realization near XLSSC~122 plays a negligible role in the noise budget of the ACA+ALMA observations ($<1~\rm \mu K$) when compared to the instrumental noise and calibration uncertainty of ALMA and ACA.} Turbulence in the atmosphere, which produces a contaminating signal for ground-based CMB observations, as well as instrumental noise, are additional sources of covariance between some combinations of the ACT maps. To account for these sources of covariance, we estimate inter-map covariance matrices using similar procedures to \citet{madhavacheril2020atacama}; we refer the reader to this paper and the references therein for a detailed description. In brief, the covariance matrix is constructed by adding a signal covariance \textit{S} and a noise covariance \textit{N}. Both of these components are obtained empirically, as follows. The noise covariance comes from nonstatic sources in time, like the instrument and the atmosphere. It is calculated by taking the difference of various splits of the data that are interleaved in time so that the static celestial signal is removed, leaving only the nonstatic noise terms behind. The signal covariance includes the contributions from components on the sky not included in our cluster model, including the CMB. Both the signal and noise covariances are estimated from $8$ patches of the sky adjacent to the field containing XLSSC~122. We assume that both the signal and noise covariances are stationary such that we can calculate the ACT likelihood in Fourier space where the covariance matrices are diagonal.

         Furthermore, in our modeling, we mainly use flat priors except for the calibration and redshift uncertainty parameters which we marginalize over. Table \ref{tab:priors} in the Appendix provides an overview of the priors used in our modeling. As we use a nested sampling implementation, we explore the entire prior volume. Hence, we compute the Bayesian evidence with a simple quadratic integration scheme using trapezoids over the initial samples \citep[see][ for more details]{Speagle2020Dynesty}, thus enabling a tool for fair model comparison. Sampling of the posterior distribution continues until the log difference of the Bayesian evidence is less than an arbitrary threshold, in our case set to $\Delta\ln\mathcal{B}=0.01$. In this work, we normalize the Bayesian evidence by the evidence of a null model to compute the Bayes factor $\mathcal{Z}$. The null evidence is computed by estimating the Gaussian likelihood when the model is set to zero while maintaining the same prior volume. When assuming a multivariate normal distribution for the posterior probability distribution, the Bayes factor can be expressed in terms of $\sigma$ through\footnote{Assuming that the posterior probability distribution follows a multivariate normal distribution, the process of likelihood marginalization needed to derive the Bayesian evidence is analogous to calculating the cumulative probability function of the underlying normal distribution. When flat priors are applied, this can be readily adjusted by introducing an analytical truncation to the likelihood function. Consequently, it becomes possible to represent the Bayesian evidence as a linear combination of scaled error functions.}
         
        \begin{equation}\label{eq:bayesian}
            \sigma = \textnormal{sgn}\left(\Delta \ln \mathcal{Z}\right) \sqrt{2\times |\Delta \ln \mathcal{Z}|}~. 
        \end{equation}
         
    \section{Interloper and cluster member removal}\label{sec:interloper_method}
    
        As the SZ effect in our observations is manifested as a decrement, we must account for any emission that would infill the signal. Such emission could arise from background sources, cluster members, and foreground interlopers such as active galactic nuclei (AGNs) or dusty star-forming galaxies observed at lower redshifts \citep[see e.g.,][]{Sayers2013, Sayers2019, Dicker2021}. Hence, we simultaneously model the extended SZ effect jointly with the more compact sources. Unresolved, compact sources are modeled using point-like (Dirac delta) emission models with a first-order polynomial to describe their spectral behavior. Often, the contaminating source is orders of magnitude brighter than the extended signal from the ICM (unlike the case for XLSSC~122). Therefore, the characterization of the ICM (for instance, the inner slope, $\gamma$ in Eq. \eqref{eq:gNFW}) could depend on how the unwanted continuum emission is subtracted from the data. By modeling both the SZ effect and contaminating sources, we have a better handle on removing contamination through Bayesian inference.

    % \subsection{Mitigating point-source contamination in the ACA and ALMA observations}\label{sec:pointsources}

            \begin{figure}[t]
                \centering
                \includegraphics[width=\hsize]{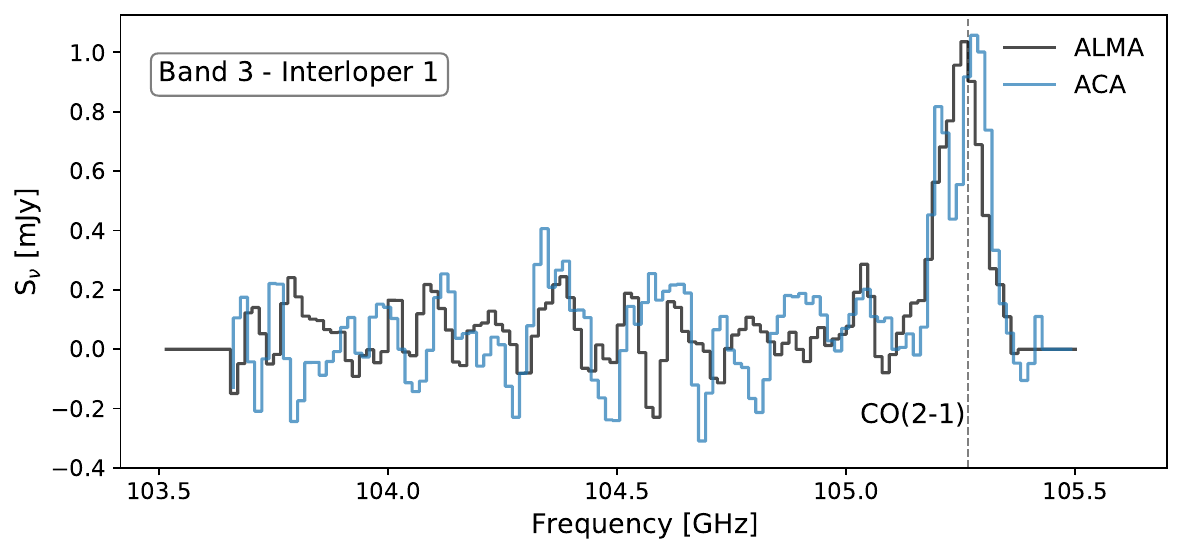}
                \includegraphics[width=\hsize]{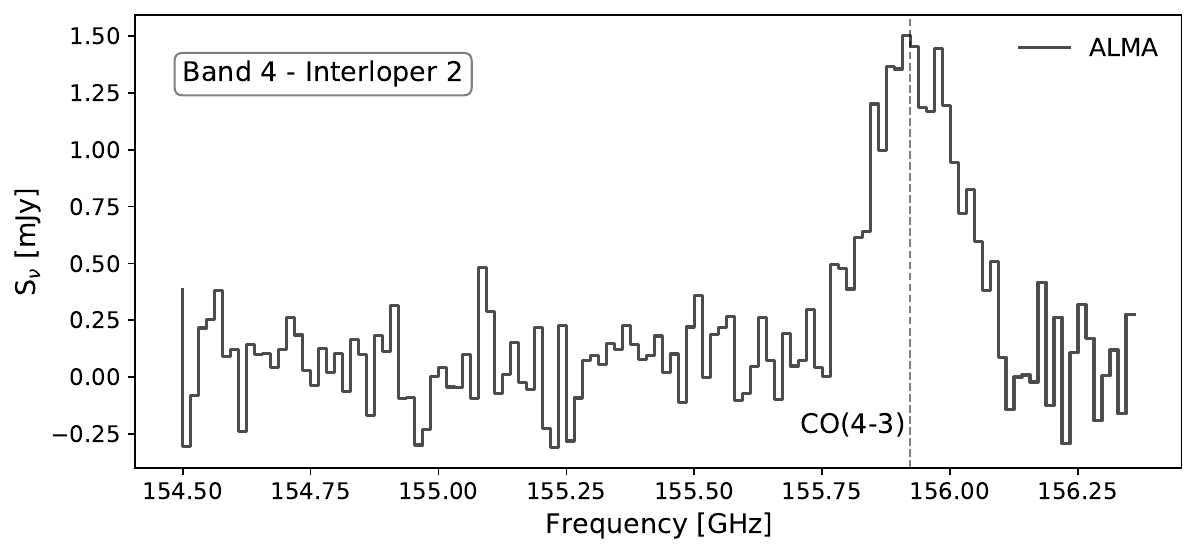}
                \caption{Emission lines in the ALMA and ACA observations obtained from a single beam. One in Band 3 and the other in Band 4 at (RA, Dec) coordinates (\ra{2;17;42.8101},\dec{-3;45;31.062}) and (\ra{2;17;41.2573}, \dec{-3;45;31.799}) respectively. The line emission comes from two galaxies at $z=1.19$ and z=$1.96$ which are also detected in optical broadband images.}
                \label{fig:strong_lines}
            \end{figure}
        
        At two locations in the ALMA and ACA image, we find evidence for bright spectral line emission, as shown in Figure~\ref{fig:strong_lines}. One line is found in Band 3 and the other in Band 4 at an RA and Dec (\ra{2;17;42.8101},\dec{-3;45;31.062}) and (\ra{2;17;41.2573},\dec{-3;45;31.799}) respectively. The lines correspond to two galaxies at $z=1.19$ and z=$1.96$ and are co-spatial with galaxies detected in the optical broad-band images. The galaxies are bright enough to show as point sources in the continuum maps (e.g., see Fig.~\ref{fig:combined}). Hence, we removed the higher-frequency half of the corresponding spectral window before continuum modeling the pressure distribution of the ICM.

        Furthermore, we detect line emission from two cluster members by visually inspecting the channel maps at the location of their optical counterparts. These sources, however, are too faint to be detected in the continuum image.\footnote{We perform additional binning along the frequency axes into a single-frequency bin to decrease computation time while minimizing the \textit{uv}-coverage loss.} Hence, these lines represent a negligible contribution to the modeling. Finally, we find no radio source in surveys such as the Very Large Array Sky Survey (VLASS, \citealt{2020PASP..132c5001L}) within our field of view, including the brightest cluster galaxy (BCG).

        With the removal of the bright emission lines, we find no further $>4\sigma$-bright point sources in the ALMA+ACA Band 3 continuum maps (e.g., see Fig.~\ref{fig:multiwavelength}). However, we find a $21\sigma$ point source continuum detection in the ALMA Band 4 observations at the location of the bright emission line seen in the Band 3 observations (\ra{2;17;42.8101},\dec{-3;45;31.062}). 
        Thus, we utilize Band 4 data to broaden the spectral coverage and remove the dust-continuum emission originating from this interfering source. We did this by modeling the point source using both ALMA Band 3 and 4 observations. We note that this point source is located just beyond $r_{500}$ and is undetected ($<2\sigma$) in the ACA+ALMA Band 3 map. Nevertheless, we subtracted the model from the observations before modeling the SZ signal, which is described in the next section.

        Finally, we ran a point source search with our forward modeling routine using both the ALMA+ACA Band 3 and ALMA Band 4 observations with a lower \textit{uv}-cut of 11~k$\lambda$ to exclude extended emission. We model any possible emission features with a point source which is described by seven parameters, namely: RA, Dec, Amplitude, spectral slope, and the three calibration uncertainty parameters of the observations. Other than the earlier mentioned two bright sources, we found no other region with a significant detection in the posterior distribution of the modeled parameters, indicating that no further point source contamination is present in the data set, including the sources for which we found tentative line emission in the channel maps. So, by adding the high-resolution ALMA+ACA observations, we find that XLSSC~122 is not significantly contaminated. The only contaminating source present is off-center and can be considered point-like.

    \section{Single component ICM modeling}\label{sec:results}

        Here we present our general results regarding the pressure profile modeling. Our baseline assumption is that the surface brightness distribution is well-described by one component, and hence the pressure distribution of the ICM is modeled with a single profile. For this, we will model using both the empirical and theoretical formalism. To assess the impact of adding data from a single-dish telescope to the forward-modeling routine, we first model the ICM using ALMA and ACA observations only. These results are given in Section~\ref{sec:single_acaalma}. The results regarding the joint modeling to the ALMA+ACA+ACT observations are provided in Section~\ref{sec:ACT-modeling}. The latter section also discusses how adding zerospacing information changes the derived posterior distributions of the modeled parameters.
        
        \begin{table*}[]
                \resizebox{\textwidth}{!}
                {%
                \begin{tabular}{lll|lllllllll}
                    \multicolumn{11}{c}{\textbf{ALMA+ACA: Single theoretical-model posterior values}} \\ 
                    \toprule
                    Model Type  & $|\Delta$ ln$\mathcal{Z}$| &   $\sigma_{\rm eff}^\dagger$            & $\Delta$RA                                   & $\Delta$Dec                                    & $P_0$                               & $r_{\rm s}$                             & e                               & PA                          & $\alpha$                     & $\beta$                       & $\gamma$                            \\ 
                    --          &    --          &     --          &    [$\arcsec$]                        & [$\arcsec$]                             & [keV~cm$^{-3}$]                    & [$^\circ$]                        &  -                              & [$^\circ$]                  & --                           & --                            & --                                  \\ 
                    \midrule
                    gNFW       &  59.9          &      10.9         &  $-0.5\substack{+0.9\\-0.9}$ & $11.0\substack{+1.2 \\ -1.2}$ & $0.07\substack{+0.02 \\ -0.02}$ & $0.011\substack{+0.003 \\-0.002}$ & 0.00                            & ~~~0.0                        & 1.0510                       & 5.4905                        & ~~~0.3081                            \\
                    gNFW       &  59.8          &      10.9         &  $-0.9\substack{+1.0\\-1.0}$ & $10.7\substack{+1.2 \\ -1.2}$ & $0.48\substack{+0.31 \\ -0.29}$ & $0.007\substack{+0.002 \\-0.001}$ & 0.00                            & ~~~0.0                        & 1.0510                       & 5.4905                        & $-0.51\substack{+0.44 \\ -0.29}$   \\
                    gNFW       &  59.5          &      10.9         &  $-0.9\substack{+0.9\\-1.0}$ & $10.7\substack{+1.2 \\ -1.2}$ & $0.43\substack{+0.34 \\ -0.26}$ & $0.010\substack{+0.004 \\-0.004}$ & 0.00                            & ~~~0.0                        & 1.0510                       & $7.1\substack{+1.9 \\ -2.5}$  & $-0.44\substack{+0.42 \\ -0.30}$   \\
                    gNFW       &  57.7          &      10.7         &  $-0.8\substack{+1.0\\-1.0}$ & $10.7\substack{+1.4 \\ -1.2}$ & $0.18\substack{+0.38 \\ -0.13}$ & $0.008\substack{+0.008 \\-0.004}$ & 0.00                            & ~~~0.0                        & $1.3\substack{+2.1 \\ -0.4}$ & $5.8\substack{+2.5 \\ -3.0}$  & $-0.17\substack{+0.53 \\ -0.42}$   \\ 
                    
                    gNFW       &  62.0          &      11.1         &  $-0.4\substack{+0.8\\-0.8}$ & $10.9\substack{+1.3 \\ -1.3}$ & $0.11\substack{+0.04 \\ -0.03}$ & $0.014\substack{+0.003 \\-0.002}$ & $0.52\substack{+0.12 \\ -0.15}$ & $-0.0\substack{+10 \\ -10}$ & 1.0510                        & 5.4905                         & ~~~0.3081                          \\
                    gNFW       &  60.7          &      11.0         &  $-0.8\substack{+0.9\\-1.0}$ & $10.7\substack{+1.3 \\ -1.4}$ & $0.39\substack{+0.36 \\ -0.22}$ & $0.009\substack{+0.003 \\-0.002}$ & $0.51\substack{+0.13 \\ -0.16}$ & $-1.0\substack{+11 \\ -11}$ & 1.0510                        & 5.4905                         & $-0.25\substack{+0.39 \\ -0.35}$ \\
                    gNFW       &  60.8          &      11.0         &  $-0.8\substack{+0.9\\-1.0}$ & $10.6\substack{+1.4 \\ -1.3}$ & $0.42\substack{+0.34 \\ -0.24}$ & $0.013\substack{+0.005 \\-0.005}$ & $0.50\substack{+0.12 \\ -0.16}$ & $-1.5\substack{+11 \\ -11}$ & 1.0510                        & $7.2\substack{+1.8 \\ -2.4}$   & $-0.28\substack{+0.37 \\ -0.31}$ \\
                    gNFW       &  59.5          &      10.9         &  $-0.9\substack{+0.9\\-0.9}$ & $10.5\substack{+1.3 \\ -1.4}$ & $0.17\substack{+0.28 \\ -0.10}$ & $0.007\substack{+0.007 \\-0.003}$ & $0.52\substack{+0.12 \\ -0.16}$ & $-3.0\substack{+11 \\ -11}$ & $1.8\substack{+3.5 \\ -0.80}$ & $5.1\substack{+2.9 \\ -2.7}$   & $-0.11\substack{+0.56 \\ -0.48}$ \\ 
                    \bottomrule 
                \end{tabular}}
                \centering
                \vspace{5mm}
                
                \resizebox{\textwidth}{!}
                {%
                \begin{tabular}{lll|llllllll}
                    \multicolumn{10}{c}{\textbf{ALMA+ACA: Single empirical-model posterior values}} \\ 
                    \toprule
                      Model Type                   & $|\Delta$ ln$\mathcal{Z}$| &    $\sigma_{\rm eff}^{\dagger}$        & $\Delta$RA                                    & $\Delta$Dec                                    & log $M_{500,\mathrm{c}}$                              & e                               & PA                          &$\alpha$ & $\beta$ & $\gamma$  \\ 
                     --                            & --             &      --      & [$\arcsec$]                            & [$\arcsec$]                             & [M$_{\odot}$]                    & --                              & [$^\circ$]                  & --       & --      & --       \\ 
                    \midrule
                    A10-UPP                        & 62.3           &    11.2        & $-0.3\substack{+0.8 \\ -0.9}$ & $11.0\substack{+1.3 \\ -1.2}$  & $14.01\substack{+0.03 \\ -0.03}$ & 0.00                            & ~~~0.00                       & 1.0510   & 5.4905   & 0.3081  \\
                    A10-MD                         & 62.6           &    11.2        & $-0.8\substack{+1.0 \\ -1.0}$ & $11.0\substack{+1.3 \\ -1.3}$  & $14.00\substack{+0.04 \\ -0.04}$ & 0.00                            & ~~~0.00                       & 1.4063   & 5.4905   & 0.3798  \\
                    A10-CC                         & 59.5           &    10.9        & $~~~0.2\substack{+0.9 \\ -0.9}$ & $10.9\substack{+1.5 \\ -1.4}$  & $13.87\substack{+0.03 \\ -0.04}$ & 0.00                            & ~~~0.00                       & 1.2223   & 5.4905   & 0.7736  \\
                    \rowcolor{gray!10} M14-UPP     & 62.5           &    11.2        & $-1.1\substack{+1.0 \\ -1.0}$ & $10.7\substack{+1.3 \\ -1.3}$  & $14.12\substack{+0.04 \\ -0.04}$ & 0.00                            & ~~~0.00                       & 2.2700   & 3.4800   & 0.1500  \\
                    \rowcolor{gray!10} M14-MD      & 62.3           &    11.2        & $-1.2\substack{+1.0 \\ -1.0}$ & $10.9\substack{+1.4 \\ -1.3}$  & $14.10\substack{+0.04 \\ -0.04}$ & 0.00                            & ~~~0.00                       & 1.7000   & 5.7400   & 0.0500  \\
                    \rowcolor{gray!10} M14-CC      & 62.7           &    11.2        & $-0.9\substack{+1.0 \\ -1.0}$ & $10.7\substack{+1.3 \\ -1.3}$  & $14.11\substack{+0.04 \\ -0.04}$ & 0.00                            & ~~~0.00                       & 2.3000   & 3.3400   & 0.2100  \\ 
                    A10-UPP                        & 63.8           &    11.3        & $-0.3\substack{+0.8 \\ -0.8}$ & $10.9\substack{+1.2 \\ -1.3}$  & $14.17\substack{+0.10 \\ -0.09}$ & $0.49\substack{+0.14 \\ -0.18}$ & $-1.6\substack{+18 \\ -12}$ & 1.0510   & 5.4905   & 0.3081  \\
                    A10-MD                         & 64.3           &    11.3        & $-0.8\substack{+0.9 \\ -0.9}$ & $10.9\substack{+1.4 \\ -1.4}$  & $14.17\substack{+0.10 \\ -0.09}$ & $0.47\substack{+0.13 \\ -0.17}$ & $~~~0.4\substack{+16 \\ -17}$ & 1.4063   & 5.4905   & 0.3798 \\
                    A10-CC                         & 61.5           &    11.1        & $~~~0.2\substack{+0.7 \\ -0.7}$ & $11.0\substack{+1.3 \\ -1.3}$  & $14.13\substack{+0.12 \\ -0.12}$ & $0.57\substack{+0.12 \\ -0.17}$ & $-4.3\substack{+10 \\ -17}$ & 1.2223   & 5.4905   & 0.7736  \\
                    \rowcolor{gray!10} M14-UPP     & 64.1           &    11.3        & $-1.1\substack{+0.9 \\ -0.9}$ & $10.6\substack{+1.3 \\ -1.4}$  & $14.30\substack{+0.10 \\ -0.09}$ & $0.47\substack{+0.13 \\ -0.16}$ & $-1.3\substack{+11 \\ -11}$ & 2.2700   & 3.4800   & 0.1500  \\
                    \rowcolor{gray!10} M14-MD      & 64.0           &    11.3        & $-1.3\substack{+0.9 \\ -0.9}$ & $10.7\substack{+1.4 \\ -1.4}$  & $14.29\substack{+0.10 \\ -0.09}$ & $0.48\substack{+0.12 \\ -0.16}$ & $-0.8\substack{+10 \\ -10}$ & 1.7000   & 5.7400   & 0.0500  \\
                    \rowcolor{gray!10} M14-CC      & 64.4           &    11.4        & $-1.0\substack{+0.9 \\ -0.9}$ & $10.5\substack{+1.3 \\ -1.3}$  & $14.29\substack{+0.09 \\ -0.09}$ & $0.48\substack{+0.12 \\ -0.17}$ & $-2.1\substack{+11 \\ -11}$ & 2.3000   & 3.3400   & 0.2100  \\ 
                    \bottomrule  
                \end{tabular}}
                \centering
                \caption{Most likely parameters for a single SZ component, modeled with a gNFW (upper) and the empirical formalism (lower). Every row corresponds to a unique run in which we varied the parameters that are listed with uncertainties. The uncertainties on the derived parameters are given as the 16th and 84th quantiles. Corner plots of these runs are shown in the supplementary material. The coordinate centers $\Delta$RA and $\Delta$Dec are given with respect to the BCG, which is located at an RA and Dec of \ra{2;17;44.2128}, \dec{-3;45;31.68}. $^\dagger$ The effective significance is computed via Eq.~\eqref{eq:bayesian}.}
                \label{table:sigle}
            \end{table*}   
    
            \begin{figure*}[t]
                \vspace{-4mm}
                \centering
                \includegraphics[width=0.99\textwidth]{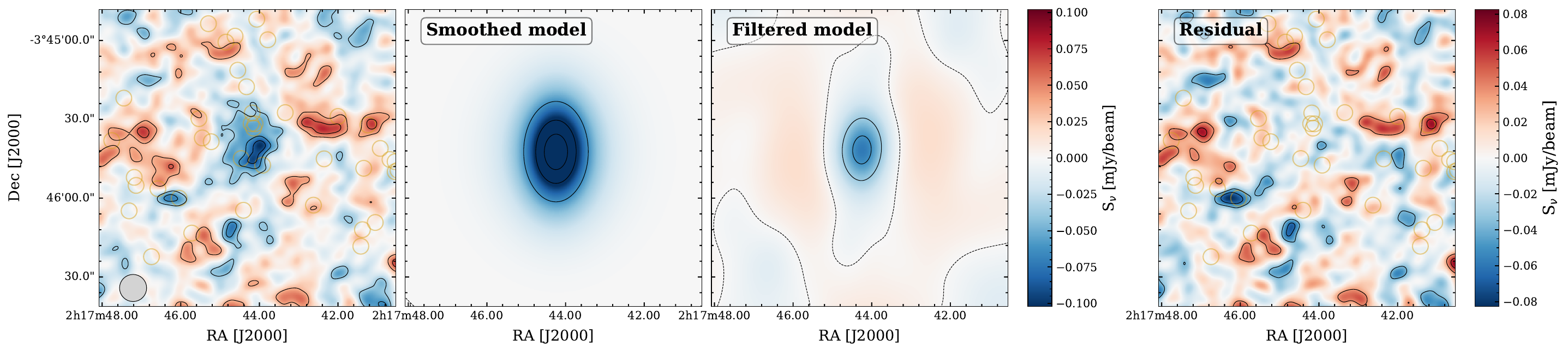}
                \caption{Most likely model from the theoretical formalism (based on the Bayes factor, row 5) of Table \ref{table:sigle}. In the left panel, we show the dirty image of the joined ACA+ALMA observations. The second panel is the likelihood-weighted model reconstruction from the nested sampling routine. The third panel shows the model corrected for the incomplete \textit{uv}-coverage of the observations. On the right, we show the residuals by subtracting the model from the observed visibilities. The black contours are drawn at [$-4.5$, $-3.5$, $-2.5$, $-1.5$, $1.5$, $2.5$, 3.5]$-\sigma$ based on the noise in the residual map. The third panel also includes the $0\sigma$ level. The contours in the second panel are drawn at [$-6.5$, $-4.5$, $-2.5$, $2.5$, $3.5$]$-\sigma$. The model shows a clear resemblance to the data.}
                \label{fig:best_image}
            \end{figure*}
        
        \subsection{ALMA+ACA modeling}\label{sec:single_acaalma}

            Table \ref{table:sigle} presents the most likely parameters derived from the nested sampling routine for both formalisms. Because of the complex degeneracies between the different model parameters (especially in the case of the theoretical model), the error on the derived parameters is asymmetric. The posterior distributions of all model runs are shown in the supplementary material. In Table \ref{table:sigle}, the parameters which are unfrozen and hence modeled are indicated with error bars. The frozen parameters are set to the shape parameters for the A10 universal pressure profile (A10-UPP) in the case of the theoretical formalism. For the empirical formalism runs, we froze the shape parameters to the averaged values found for the three different cluster classifications in \citet{Arnaud2010} and \citet{McDonald2014}. The results on the theoretical formalism and the empirical formalism are provided in Sections \ref{sec:gnfw_single}~\&~\ref{sec:empirical_single}, respectively. For all runs, we marginalized over the calibration and redshift uncertainties. All priors are given in the Appendix, Table~\ref{tab:priors}.
            
            The second column in Table~\ref{table:sigle} shows the Bayes factor $\mathcal{Z}$ which is the Bayesian evidence normalized by the evidence for the null model. Using Equation~\eqref{eq:bayesian}, we can conclude that we detect the SZ effect at $\sim10.9-11.2\sigma$ in the ACA+ALMA data, depending on the model. In the next two sections, we will go into more depth on the results shown in Table~\ref{table:sigle}.
            
        \subsubsection{The theoretical formalism}\label{sec:gnfw_single}

            \begin{figure*}[t] 
                \centering
                \includegraphics[width=0.47\textwidth]{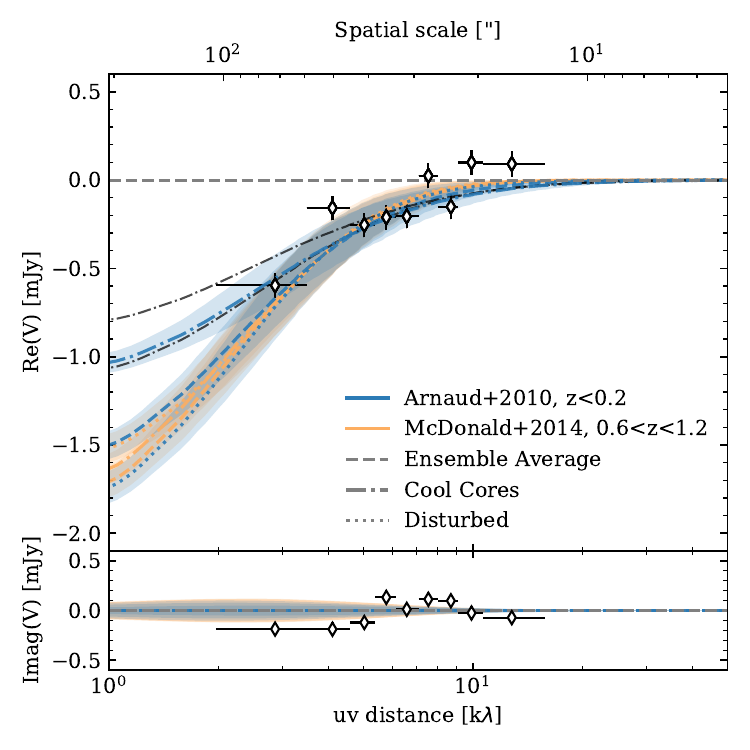}
                \includegraphics[width=0.47\textwidth]{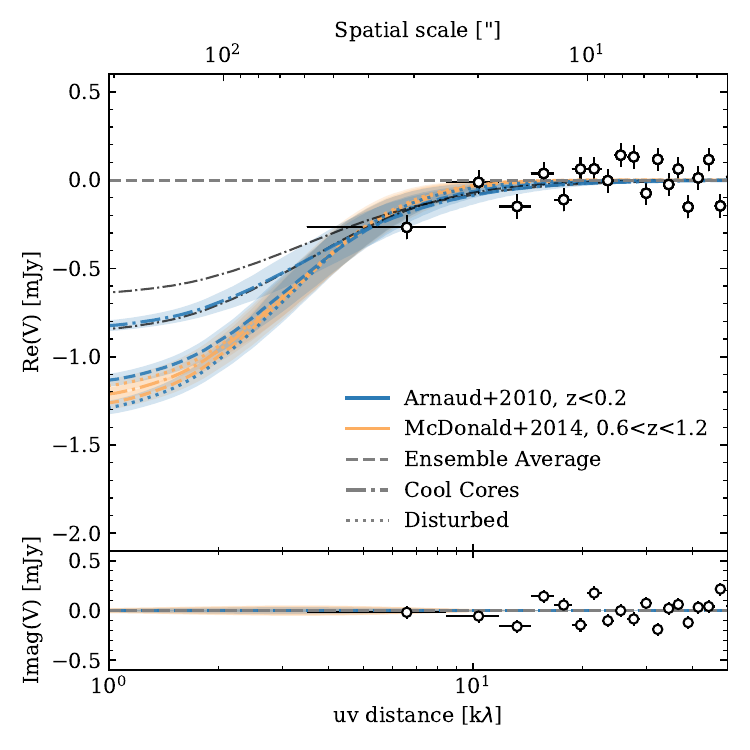}
                \caption{\textit{uv}-radial distributions of the Band 3 ACA (left) and ALMA (right) observations in which the visibilities are phase-shifted to the center of the SZ decrement. The colored lines show the median primary beam-attenuated elliptical SZ models based on the empirical formalism which are reported in Table~\ref{table:sigle}. The shaded regions indicate the standard deviation in the \textit{uv}-radial bin of the model and are thus a direct consequence of the eccentricity of the cluster. To gauge the uncertainty, we show the 0.16-0.84 quantiles of the mean A10-cc profile with the gray dash-dotted lines. The divergence of the A10-CC profile from the rest of the models and the introduction of eccentricity to the modeling of the SZ signal shows that the data disfavors a local cool core cluster as the morphological state of XLSSC~122. Other classifications cannot be separated because of the limited capabilities of ALMA+ACA in constraining fluxes at scales larger than about $2\farcm0$ ($\sim1~k\lambda$).}
                \label{fig:uv-radial}
            \end{figure*}
                   
            \begin{figure}[t]
                \centering
                \includegraphics[width = \hsize]{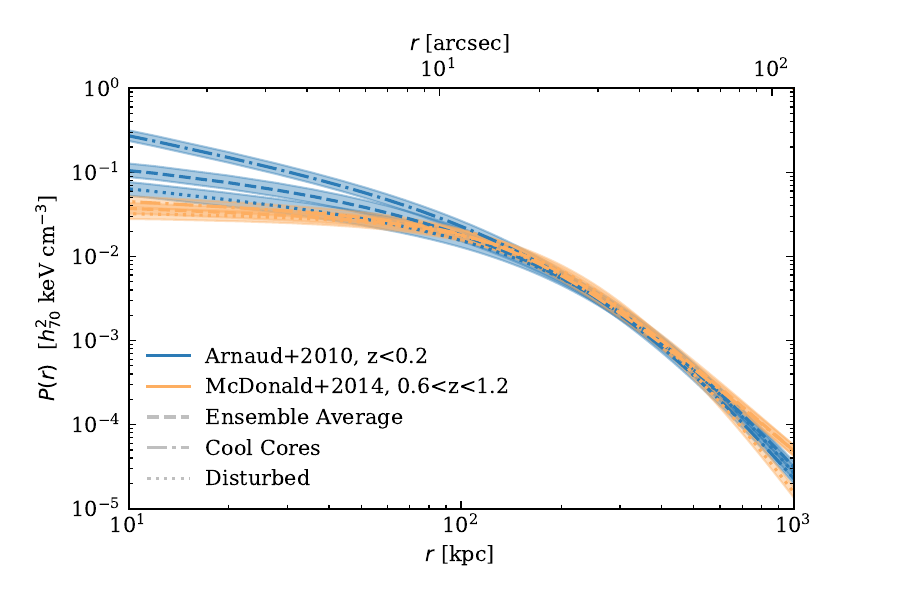}
                \caption{Derived radial pressure profiles of the best-fit elliptical models which follow the empirical formalism as shown in Table~\ref{table:sigle}. The shape parameters of the A10-model parameters are frozen and set to the six classifications of \citet{Arnaud2010} and \citet{McDonald2014}. The uncertainties are derived from the 0.16-0.84 quantiles of the sampled posterior distribution's mass (amplitude) parameter. The uncertainties are thus marginalized over the centroid position, eccentricity, redshift, and calibration uncertainties. These profiles are the image plane variants (Fourier transforms) of the ones shown in Figure~\ref{fig:uv-radial}.}
                \label{fig:pressure}
            \end{figure}

            When modeling with the theoretical formalism (shown at the top of Table~\ref{table:sigle}), we unfreeze the shape parameters one by one, thus systematically increasing the prior volume; as we do so, $|\Delta\ln \mathcal{Z}|$ systematically decreases (rows 1-4 in Table \ref{table:sigle}). Thus we conclude, via the Bayesian evidence, that we cannot robustly differentiate between runs that are modeled with more complexity (more unfrozen parameters) and the more simplistic ones because of the loss of large-scale information in the interferometric observations. This filtering removes the spatial scales which are sensitive to the total flux of the system and to the shape of the pressure profile in the outer regions, namely the $\beta$-parameter.\footnote{We note that this is not due to a radial dependence of the filtering effect but rather due to the fact that the radial trend of the pressure profile in cluster outskirts is generally too shallow (i.e., described mostly by large-scale modes). Any small-scale feature in the outer regions with an extent matching the range of scales probed by ALMA would still be observed.} As the integrated pressure along the line of sight is degenerate with $\beta$, it becomes difficult to constrain the normalization of the profile and thus fit for $P$ when $\beta$ is unfrozen. This is especially clear when modeling with the empirical-model implementation described in the next section. However, in these single component fits, we do find overall higher evidence for elliptical models with an axis ratio of $\sim50\%$ elongated nearly along the north-south axis.
            
            Figure~\ref{fig:best_image} shows the most likely model, based on the Bayesian evidence shown in Table~\ref{table:sigle}, for the theoretical formalism. The model, shown in the second panel, is made by imaging each sample of the forward-modeling routine and weighting it by its likelihood when averaging them together and illustrates the high eccentricity of the system. The third panel shows the large-scale filtering by the incomplete \textit{uv}-coverage on the images. The residuals in Figure \ref{fig:best_image} are obtained by subtracting the model in the visibility plane from the data. As the residuals indicate a good agreement between the observations and the modeled surface brightness, the base assumption that the bulk of the ICM is composed of a single pressure profile thus holds for this $z\sim2$ cluster.

        \begin{table*}[t]
            \resizebox{\textwidth}{!}
            {%
            \begin{tabular}{lll|lllllllll}
                \multicolumn{11}{c}{\textbf{ALMA+ACA+ACT: Single theoretical-model posterior values}} \\ 
                \toprule
                Model Type  & $|\Delta$ ln$\mathcal{Z}$| & $\sigma_{\rm eff}$     & $\Delta$RA                           & $\Delta$Dec                            & $P_0$                               & $r_{\rm s}$                             & e                               & PA                          & $\alpha$                     & $\beta$                         & $\gamma$                            \\ 
                --          &    --                      &  --                    &   [$\arcsec$]                        & [$\arcsec$]                            & [keV~cm$^{-3}$]                     & [$^\circ$]                              &  -                              & [$^\circ$]                  & --                           & --                              & --                                  \\ 
                \midrule
                gNFW        &  117.2                     &  15.3                  &  $-0.2\substack{+0.9\\-0.9}$        & $11.5\substack{+1.2 \\ -1.2}$          & $0.05\substack{+0.01 \\ -0.01}$     & $0.017\substack{+0.002 \\-0.001}$       & 0.00                            & ~0.0                        & 1.0510                       & 5.4905                          & ~~~0.3081                            \\
                gNFW        &  115.5                     &  15.2                  &  $-0.1\substack{+0.9\\-1.0}$         & $11.5\substack{+1.3 \\ -1.3}$          & $0.04\substack{+0.07 \\ -0.02}$     & $0.018\substack{+0.005 \\-0.005}$       & 0.00                            & ~0.0                        & 1.0510                       & 5.4905                          & ~~~$0.41\substack{+0.26 \\ -0.39}$   \\
                gNFW        &  115.5                     &  15.2                  &  $-0.2\substack{+1.0\\-1.0}$         & $11.3\substack{+1.2 \\ -1.2}$          & $0.10\substack{+0.26 \\ -0.07}$     & $0.009\substack{+0.010 \\-0.005}$       & 0.00                            & ~0.0                        & 1.0510                       & $4.0\substack{+1.5 \\ -0.8}$    & ~~~$0.16\substack{+0.39 \\ -0.66}$   \\
                gNFW        &  114.2                     &  15.1                  &  $-0.2\substack{+1.0\\-1.0}$         & $11.5\substack{+1.2 \\ -1.2}$          & $0.34\substack{+0.53 \\ -0.25}$     & $0.002\substack{+0.008 \\-0.001}$       & 0.00                            & ~0.0                        & $0.7\substack{+0.1 \\ -0.1}$ & $5.8\substack{+1.1 \\ -1.0}$  & $-0.01\substack{+0.37 \\ -0.30}$   \\ 
             
                gNFW        &  118.8                     &  15.4                  &  $-0.3\substack{+0.9\\-0.9}$         & $11.3\substack{+1.3 \\ -1.3}$          & $0.06\substack{+0.01 \\ -0.01}$     & $0.020\substack{+0.002 \\-0.002}$       & $0.47\substack{+0.12 \\ -0.15}$ & $6\substack{+10 \\ -10}$  & 1.0510                        & 5.4905                         & ~~~0.3081                          \\
                gNFW        &  117.2                     &  15.3                  &  $-0.2\substack{+0.9\\-0.9}$         & $11.3\substack{+1.3 \\ -1.3}$          & $0.05\substack{+0.07 \\ -0.02}$     & $0.022\substack{+0.006 \\-0.006}$       & $0.49\substack{+0.11 \\ -0.14}$ & $6\substack{+10 \\ -09}$  & 1.0510                        & 5.4905                         & $~~~0.43\substack{+0.22 \\ -0.35}$ \\
                % gNFW       &  -----         &  $34.4425\substack{+0.00004\\-0.00002}$ & $-3.7684\substack{+0.00001 \\ -0.000007}$ & $0.0067\substack{+0.00004 \\ -0.00004}$ & $0.0005\substack{+0.005 \\-0.005}$ & $0.51\substack{+0.002 \\ -0.002}$ & $45\substack{+0.2 \\ -0.2}$ & 1.0510                        & $0.5\substack{+0.0004 \\ -0.0003}$   & $3.27\substack{+0.54 \\ -0.92}$ \\
                
                % gNFW       &  -----         &  $34.4425\substack{+0.00002\\-0.00004}$ & $-3.7519\substack{+0.00002 \\ -0.00004}$ & $0.015\substack{+0.0002 \\ -0.0002}$ & $0.004\substack{+0.0000 \\-0.0000}$ & $1.00\substack{+0.08 \\ -0.08}$ & $-42\substack{+11 \\ -11}$ & $3.7\substack{+3.2 \\ -2.0}$ & $0.678\substack{+0.003 \\ -0.002}$   & $-1.41\substack{+1.72 \\ -1.57}$ \\ 
                \bottomrule 
            \end{tabular}}
            \centering
            \vspace{5mm}
          
            \resizebox{\textwidth}{!}
            {%
            \begin{tabular}{lll|llllllll}
                \multicolumn{10}{c}{\textbf{ALMA+ACA+ACT: Single empirical-model posterior values}} \\ 
                \toprule
                  Model Type                   & |$\Delta$ ln $\mathcal{Z}$| &  $\sigma_{\rm eff}$    & $\Delta$RA                                    & $\Delta$Dec                                    & log $M_{500,\mathrm{c}}$                             & e                               & PA                          &$\alpha$ & $\beta$ & $\gamma$  \\ 
                 --                            & --             &   --       & [$\arcsec$]                            & [$\arcsec$]                             & [M$_{\odot}$]                    & --                              & [$^\circ$]                  & --       & --      & --       \\ 
                \midrule
                A10-UPP                        & 111.2          &  14.9    & $-0.1\substack{+0.7 \\ -0.8}$  & $10.9\substack{+1.1 \\ -1.1}$  & $14.09\substack{+0.02 \\ -0.02}$ & 0.00                            & ~0.00                       & 1.0510   & 5.4905   & 0.3081  \\
                A10-MD                         & 114.6          &  15.1    & $-0.4\substack{+0.9 \\ -0.8}$  & $11.2\substack{+1.2 \\ -1.2}$  & $14.09\substack{+0.02 \\ -0.03}$ & 0.00                            & ~0.00                       & 1.4063   & 5.4905   & 0.3798  \\
                A10-CC                         & ~~94.7         &  13.8    & $~~~0.6\substack{+0.8 \\ -0.8}$ & $10.1\substack{+1.1 \\ -1.3}$  & $13.98\substack{+0.03 \\ -0.03}$ & 0.00                            & ~0.00                       & 1.2223   & 5.4905   & 0.7736  \\
                \rowcolor{gray!10} M14-UPP     & 114.9          &  15.2    & $-0.8\substack{+0.9 \\ -0.9}$  & $10.9\substack{+1.2 \\ -1.1}$  & $14.21\substack{+0.02 \\ -0.03}$ & 0.00                            & ~0.00                       & 2.2700   & 3.4800   & 0.1500  \\
                \rowcolor{gray!10} M14-MD      & 109.9          &  14.8    & $-1.0\substack{+0.9 \\ -0.9}$  & $11.2\substack{+1.2 \\ -1.3}$  & $14.21\substack{+0.03 \\ -0.03}$ & 0.00                            & ~0.00                       & 1.7000   & 5.7400   & 0.0500  \\
                \rowcolor{gray!10} M14-CC      & 114.0          &  15.1    & $-0.6\substack{+0.9 \\ -0.9}$  & $10.7\substack{+1.1 \\ -1.2}$  & $14.21\substack{+0.03 \\ -0.03}$ & 0.00                            & ~0.00                       & 2.3000   & 3.3400   & 0.2100 \\ 
                A10-UPP                        & 110.9          &  14.9    & $-0.0\substack{+0.8 \\ -0.7}$  & $10.7\substack{+1.1 \\ -1.2}$  & $14.18\substack{+0.08 \\ -0.06}$ & $0.32\substack{+0.18 \\ -0.19}$ & $-2\substack{+13 \\ -14}$ & 1.0510   & 5.4905   & 0.3081  \\
                A10-MD                         & 115.7          &  15.2    & $-0.4\substack{+0.8 \\ -0.7}$  & $11.0\substack{+1.2 \\ -1.2}$  & $14.24\substack{+0.08 \\ -0.08}$ & $0.42\substack{+0.13 \\ -0.17}$ & $~~~1\substack{+10  \\ -10}$ & 1.4063   & 5.4905   & 0.3798\\
                A10-CC                         & ~~99.7         &  14.1    & $~~~0.7\substack{+0.6 \\ -0.7}$ & $10.4\substack{+1.3 \\ -1.4}$  & $14.37\substack{+0.11 \\ -0.11}$ & $0.71\substack{+0.07 \\ -0.10}$ & $-3\substack{+5 \\ -5}$ & 1.2223   & 5.4905   & 0.7736  \\
                \rowcolor{gray!10} M14-UPP     & 116.4          &  15.3    & $-0.8\substack{+0.8 \\ -0.8}$  & $10.8\substack{+1.3 \\ -1.2}$  & $14.37\substack{+0.08 \\ -0.08}$ & $0.45\substack{+0.12 \\ -0.16}$ & $~~~0\substack{+9 \\ -9}$ & 2.2700   & 3.4800   & 0.1500  \\
                \rowcolor{gray!10} M14-MD      & 111.0          &  14.9    & $-1.0\substack{+0.8 \\ -0.9}$  & $10.9\substack{+1.2 \\ -1.3}$  & $14.35\substack{+0.08 \\ -0.07}$ & $0.42\substack{+0.13 \\ -0.16}$ & $~~~0\substack{+10 \\ -10}$ & 1.7000   & 5.7400   & 0.0500  \\
                \rowcolor{gray!10} M14-CC      & 115.8          &  15.2    & $-0.6\substack{+0.8 \\ -0.7}$  & $10.6\substack{+1.2 \\ -1.2}$  & $14.38\substack{+0.09 \\ -0.08}$ & $0.48\substack{+0.12 \\ -0.15}$ & $-1\substack{+8 \\ -9}$ & 2.3000   & 3.3400   & 0.2100  \\ 
                \bottomrule  
            \end{tabular}}
            \centering
            \caption{Similar as Table~\ref{table:sigle} but modeled with ALMA+ACA+ACT observations.}
            \label{table:sigle_ACA_ALMA_ACT}
            \vspace{-10pt}
        \end{table*} 

        \begin{figure*}[t]
                \includegraphics[width = 0.32\textwidth]{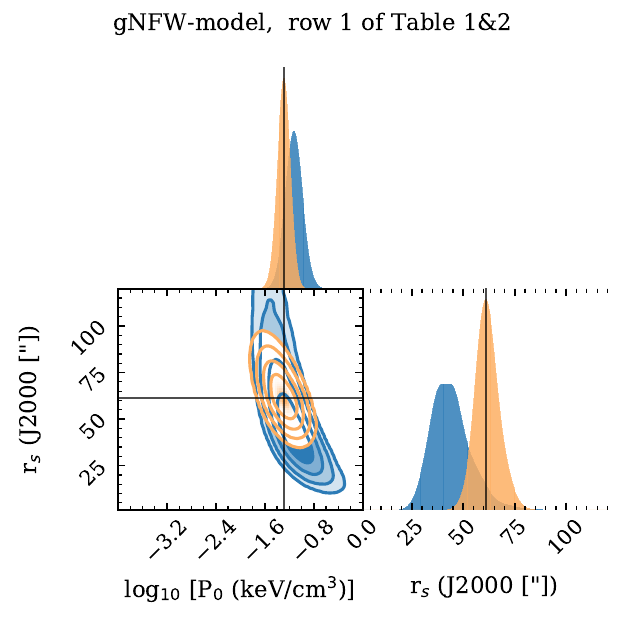}
                \includegraphics[width = 0.32\textwidth]{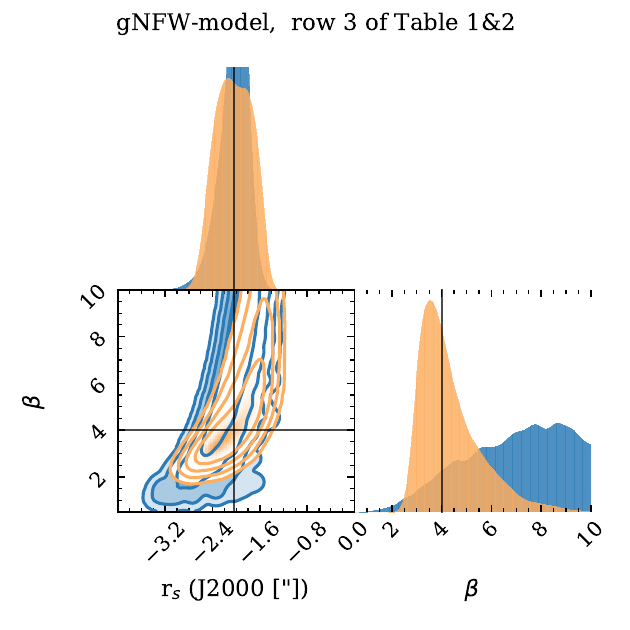} \\
            \sidecaption
                \includegraphics[width = 0.32\textwidth]{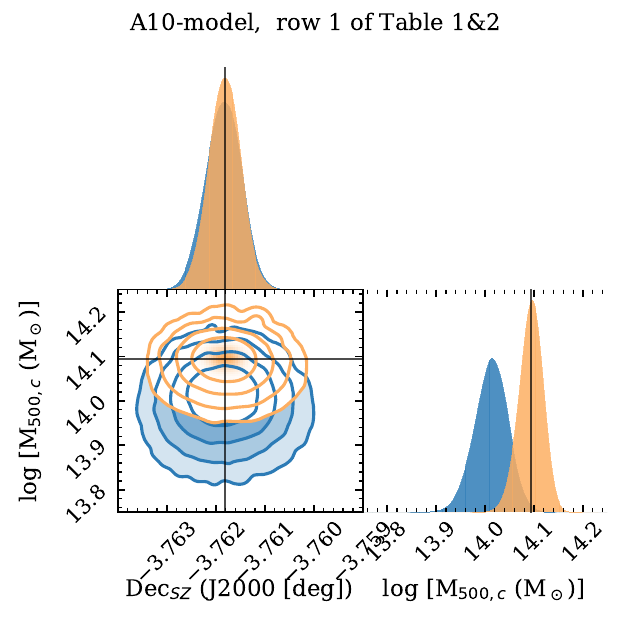} 
                \includegraphics[width = 0.32\textwidth]{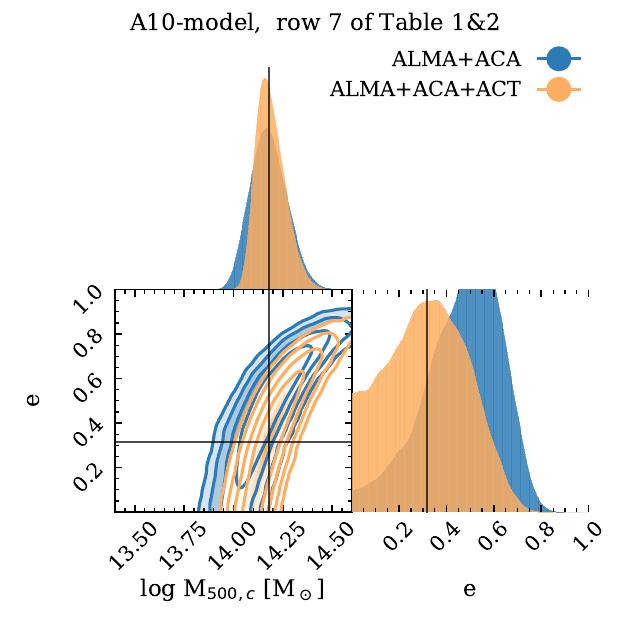}
                \caption{Comparison between modeling with ALMA+ACA (blue) and ALMA+ACA+ACT (orange) observations. The full posterior distributions of all runs can be found in the supplementary material. Here we highlight four marginalized posterior distributions to show the effect ACT observations have on the $P_0-r_s$, $r_s-\beta$, Dec$-M_{500,c}$, and the $M_{500,c}-e$ relationships, shown from left to right, top to bottom. The weighted median values of the ALMA+ACA+ACT runs are highlighted with black lines. For a detailed description of the effect the ACT observations have on the sampled posterior distribution, we refer to the text in section~\ref{sec:ACT-modeling}.}
                \label{fig:corner_highlights}
        \end{figure*}

            The modeling shows a preference for negative to flat $\gamma$ solutions, indicating a disturbance of the pressure profile in the inner regions of the cluster. However, another plausible explanation for low $\gamma$-values would be that the positive continuum emission of the central galaxies balances the negative surface brightness of the SZ decrement. From the X-ray imaging (see Fig.~\ref{fig:multiwavelength}), there is a small hint of possible AGN activity as a slight increase in emissions is found at the location of the BCG. However, we do not detect the BCG or any of the central galaxies in the ALMA and ACA data in either the continuum or any of the spectral lines (because of the spectral tuning of the observations). However, Figure~\ref{fig:multiwavelength} shows a decrease in the decrement around the central four galaxies, indicating a possible presence of continuum emission from the BCG. To test this, we reran the best-fitting gNFW model (row five in Table~\ref{table:sigle}) with an additional point source component frozen at the location of the BCG (based on the HST imaging). We modeled (including the Band 4 data) both the amplitude and spectral slope of the central galaxy. We find that the amplitude of any point source at the location of the BCG is consistent with noise. Then, by resampling the parameter space through the nested sampling routine, we find no significant difference in the derived parameters of the gNFW profile, including the inner slope value, $\gamma$, when including a point source component. Through the Bayes factor, we can reject the presence of a point source at the location of the BCG in the ALMA+ACA data by $4.1\sigma$. These results are similar to the findings of \citet{2023PASJ...75..311K} on the effect of low $S/N$ contaminating sources when deriving pressure profiles from the SZ effect. Furthermore, the additional point source component did not affect the centroid position of the SZ effect. 
            
        \subsubsection{The empirical formalism}\label{sec:empirical_single}

             \begin{figure*}[t] 
                \centering
                \includegraphics[width=0.47\textwidth]{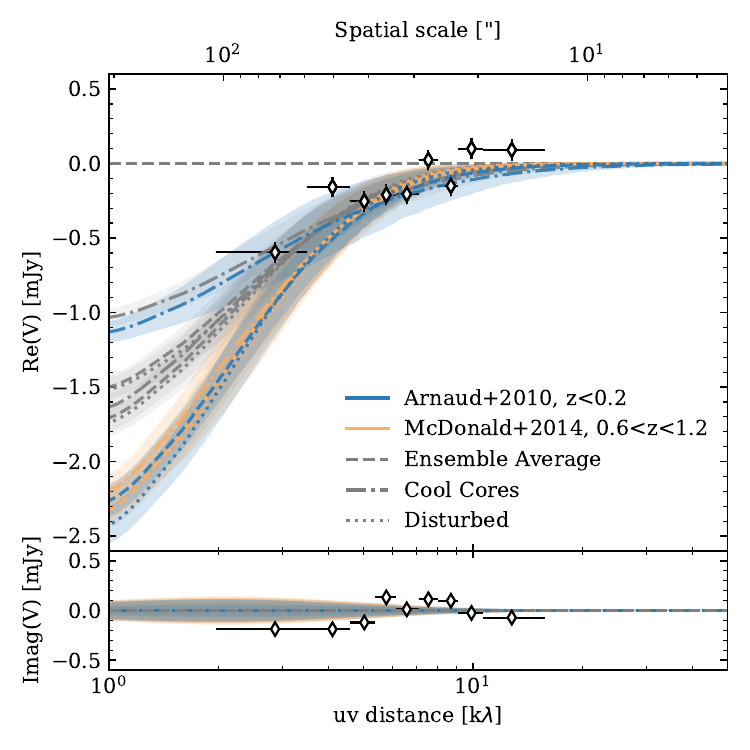}
                \includegraphics[width=0.47\textwidth]{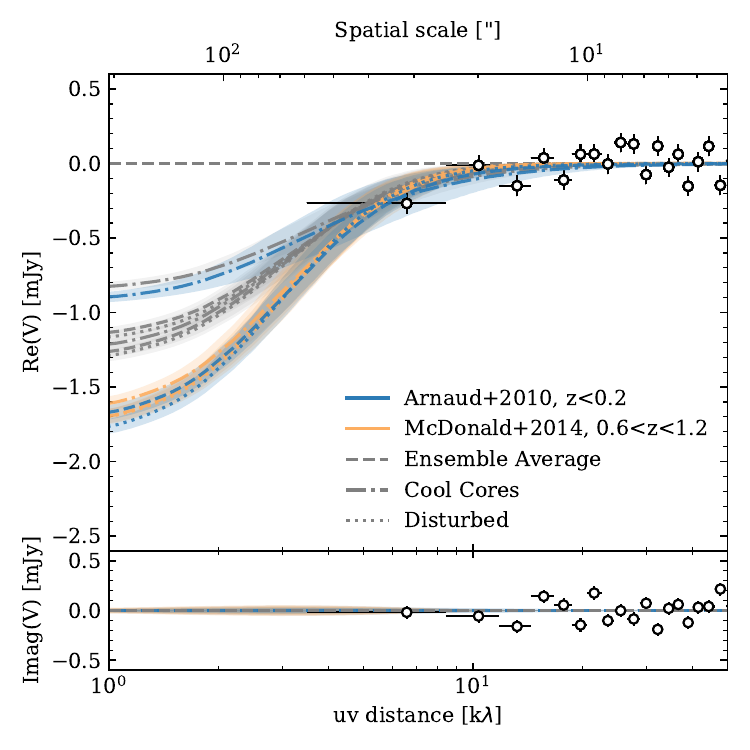}
                \caption{Same as Figure \ref{fig:uv-radial} (also shown here in gray), but including here the additional constraining power of the ACT data in the model fitting. We observe a tighter correlation at smaller \textit{uv}-distances, an overall larger amplitude, and a smaller eccentricity. We note the ACT data are not shown due to the complexity of accurately representing the image space data in this domain.}
                \label{fig:uv-radial_andact}
            \end{figure*}
            
            \begin{figure}[t]
                \centering
                \includegraphics[width = \hsize]{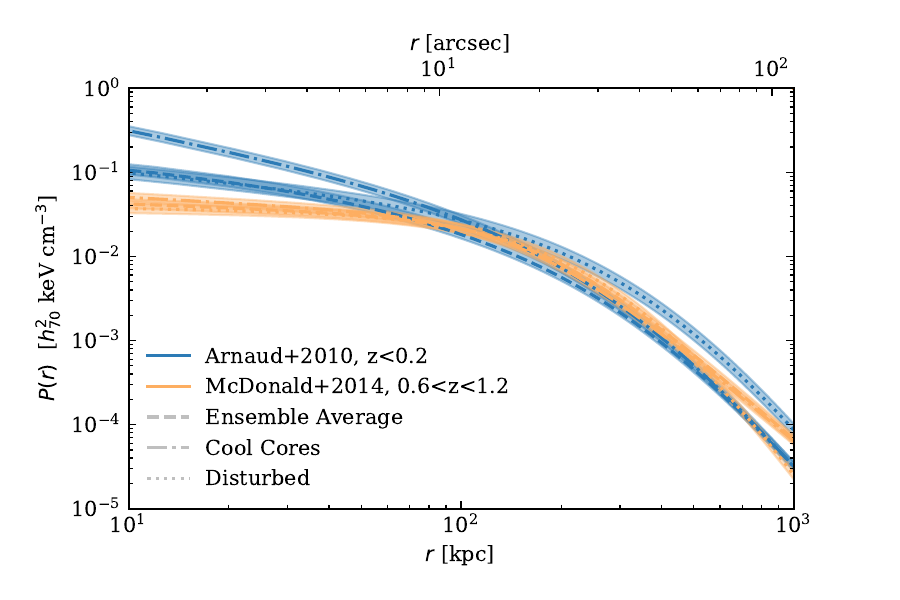}
                \caption{Same as Figure \ref{fig:pressure} but with the addition of ACT constraints, as in Figure \ref{fig:uv-radial_andact}.
                }
                \label{fig:pressure_andact}
            \end{figure}
        
            For runs with the A10 (empirical) formalism, we did not model any of the shape parameters but set them to the values for each separate classification as described in \citet{Arnaud2010} and \citet{McDonald2014}. In the modeling, we freed the centroid, eccentricity, position angle, and mass of the system, which is linked to the amplitude and scale radius of the pressure profile (see section \ref{sec:parametric}). By freeing up the amplitude parameter, the forward modeling routine automatically corrected for the unequal $Y$-value of each classification. Hence, we find for the different model classifications a different mass estimate. However, the Bayesian evidence tells us (second column in Table~\ref{table:sigle}) that the data cannot distinguish which model type is preferred when the total mass (or the total Compton-$Y$ value) is unknown. Only the pressure profile that corresponds to the averaged local cool-core (A10-CC) profile is disfavored by the data ($\Delta\ln\mathcal{Z} > 2.3$).
            
            When estimating the significance between various models, one cannot simply take the difference between two $\sigma_{\rm eff}$ values reported in the third column of Table~\ref{table:sigle}. For example, the significance between the spherical symmetric A10-MD and A10-CC is, $\sigma_{\rm diff} = \sqrt{2\times(62.3-59.5)} = 2.4$, meaning that the spherical A10-MD model is favored by the data with a significance of $2.4\sigma$ over the A10-CC one assuming a multivariate normal distribution for the posterior probability distribution, which is only a modest improvement. As the normal distribution is not a perfect fit for the posterior distribution, we can also look at the difference in the Bayes factor for the two classifications: $\Delta \ln\mathcal{Z} = 62.3-59.5 = 2.8$, which according to standard Bayesian inference is interpreted as a strong (but not definitive) evidence for preferring the A10-MD model over the A10-CC one \citep{Jeffreys1961}. This is consistent with the finding from the theoretical formalism in the previous section that a small or negative $\gamma$ solution is mildly preferred by the data.
            
           Furthermore, the lack of short baselines in the ALMA+ACA observations explains why the mass estimates vary between models and why the data do not have a significant preference among the various templates. The latter statement is a consequence of the strong $\beta$--$P_0$ degeneracy which led to the inability to constrain $\beta$ in the theoretical formalism. Figure~\ref{fig:uv-radial} shows the constraining power of the ACA and ALMA observations together with the median models from Table~\ref{table:sigle} in the Fourier domain, the domain where the likelihood is estimated. Figure~\ref{fig:uv-radial} is made by phase shifting the respective ALMA and ACA pointings to the centroid of the SZ decrement and then radially binning the $uv$-coordinates in spherically symmetric shells. The radial bins are spaced such that the statistical uncertainty in each bin is equal (i.e., each bin comprises an equal number of visibilities). The error bars show the mean and standard deviation for the real and imaginary parts of the visibilities for the stacked pointings. The ACA mosaic has no field pointing to the center of the mosaic. Therefore, we used the three ACA fields closest and with similar distances to the SZ centroid to combine the visibilities in \textit{uv}-space. Adding other fields at a larger angular separation from the SZ centroid would cause discrepancies regarding the $uv$ radial amplitude because of the primary beam attenuation of the antennas. For the ALMA observations, we show the central field in Figure~\ref{fig:uv-radial}. The same operation as done on the data is performed on the primary-beam attenuated SZ models shown in Table~\ref{table:sigle}. We note that the modeling is performed on the unbinned two-dimensional visibilities, not these radially binned ones. The shaded regions in Figure~\ref{fig:uv-radial} indicate the standard deviation in the \textit{uv}-radial bin of the model, not the error provided in the posterior distribution, and are thus a direct consequence of the eccentricity of the cluster.
           
          Figure~\ref{fig:uv-radial} makes clear that we cannot distinguish between the different classifications when modeling for the projected eccentricity of the cluster. For intuition's sake, Figure~\ref{fig:pressure} shows the image plane variant of Figure~\ref{fig:uv-radial}, namely the derived pressure profiles. Here we calculated the uncertainties based on the samples of the nested sampling routine. All in all, one needs to be careful when deriving pressure profiles when the total flux of the system is unknown. Hence, we turn in Section~\ref{sec:ACT-modeling} to provide the results of the ALMA+ACA+ACT modeling.
            
    \subsection{ALMA+ACA+ACT joint likelihood modeling}\label{sec:ACT-modeling}
 
        By adding single-dish ACT observations, we get an additional constraint on the total flux of XLSSC~122. Even though XLSSC~122 is almost unresolved in ACT, as shown in Figure~\ref{fig:combined}, we still propagate the full ICM model to the ACT frequency maps when forward modeling as described in section \ref{sec:act_method}. The results for both the theoretical and empirical formalism are shown in Table~\ref{table:sigle_ACA_ALMA_ACT}. In Figure~\ref{fig:corner_highlights}, we highlight four marginalized posterior distributions to show the effect that adding single-dish ACT observations to the forward-modeling routine has on the derived parameters. The full corner plots of the runs are included in the supplementary material. 

        Figures~\ref{fig:uv-radial_andact} and \ref{fig:pressure_andact} show the \textit{uv}-radial distributions of the modeled pressure profiles and its image-plane variant. We observe a tighter scatter at smaller \textit{uv}-distances and a smaller eccentricity (also shown in Table~\ref{table:sigle_ACA_ALMA_ACT} and the fourth panel of Fig.~\ref{fig:corner_highlights}). In the ALMA+ACA runs, the eccentricity (defined as one minus the minor over major axis ratio) was, on average, $e=0.50^{+0.12}_{-0.16}$. With the introduction of the ACT observations, we find that the eccentricity is driven to lower values with a likelihood-weighted average of %$e=0.38^{+0.14}_{-0.17}$. 
        $e=0.46^{+0.12}_{-0.16}$.
        This makes the eccentricity more consistent with 0, but regardless, Figure~\ref{fig:corner_highlights} shows that the modeling still has constraining power on $e$.
        Furthermore, the A10-CC profile model is disfavored by the data ($\Delta\ln\mathcal{Z}>11.2$) and seems to compensate for the higher mass value with a larger minor over major axis ratio. The eccentricity is modeled by squeezing the grid on which the pressure distribution is mapped to the desired extent while assuming a nonzero eccentricity along the line of sight direction. This compression leads to a reduction in the integrated pressure along the line of sight, consequently decreasing the total flux. Hence, the eccentricity is degenerate with the amplitude and $\beta$-parameter of the pressure profile creating the wide posterior distribution shown in Figure~\ref{fig:corner_highlights}. 
        
        Furthermore, with the inclusion of the ACT observations, we find higher halo mass estimates 
        % (the likelihood weighted standard deviations of the $\log M_{500,c}$ parameter in Table~\ref{table:sigle} \& \ref{table:sigle_ACA_ALMA_ACT} are $\sigma_{\log M_{\rm 500,c}} = 0.12~M_\odot$ vs $\sigma_{\log M_{\rm 500,c}} = 0.10~M_\odot$, respectively) 
        ($>3\sigma$) for the spherically symmetric models, but the significance vanishes when the eccentricity is taken into account with the modeling. We do find systematically tighter uncertainties on the halo mass (by a factor of $\sim1.5$) among the different models when modeling only to the ACA+ALMA observations. This is also shown in the third panel of Figure~\ref{fig:corner_highlights}. This panel indicates that the inclusion of the ACT observations leads to a tighter constraint on $M_{500,c}$ while the centroid of the SZ effect is unaffected by the ACT-observation and driven by the ALMA+ACA observations which probe much smaller scales. Still, the \citet{McDonald2014} profiles consistently estimate higher masses; however, the discrepancy becomes insignificant ($<2\sigma$) when the eccentricity is included in the modeling. 

        Furthermore, we still cannot significantly differentiate between the several profile classifications by adding ACT data. This is because of the high-dimensional parameter space, the complex degeneracies, the relatively low $S/N$ in the spatial scales $3-10~\mathrm{k}\lambda$ of the ACA and ALMA observations, and the similarity between the different profiles. But guided by the Bayesian evidence ($\Delta \ln{\mathcal{Z}}>4.7$), we can say that XLSSC~122 is similar to local morphologically disturbed clusters (A10-MD) and the more distant relaxed clusters (M14-UPP and M14-CC). The M14-MD model is disfavored ($\Delta \ln \mathcal{Z}> 4.8$) likely because of the relatively large $\beta$ value, $\beta = 5.74$. From the ALMA+ACA+ACT modeling with the theoretical formalism, we find consistently lower beta values which is in line with the Bayes factor disfavoring the M14-MD model over the other ones.

        Regarding the modeling based on the theoretical formalism, the first point to note is that the two gNFW runs in which we additionally unfroze the parameters $\gamma$, $\beta$ and $\gamma$, $\beta$, and $\alpha$ for elliptical morphologies both resulted in highly unphysical solutions for the best-fit model parameters. For instance, the eccentricity parameter $e$ converged to extreme values of $e>0.9$. Similarly, the posterior distribution for the $\beta$ slope saturated over the lower prior edge, corresponding to $\beta<2$ and, in turn, to the impossibility of deriving a numerical solution to the line-of-sight pressure integral. Hence, these models are not reported in Table~\ref{table:sigle_ACA_ALMA_ACT}.
        For the remaining cases, the introduction of the ACT data significantly mitigated the $P_0-r_{\rm s}$ degeneracy (see the first panel of Fig.~\ref{fig:corner_highlights}). This translated to tighter constraints on the derived shape parameters of the gNFW-formalism: $\gamma$ (the inner slope parameter) is now for all runs more consistent with zero and positive values, and the constraints on $\beta$ (the outer slope parameter) are now consistent with derived values from hybrid analysis of observations and hydrodynamical simulations \citep{Arnaud2010} and stacked \textit{Chandra} observations \citep{McDonald2014} while the corner plots from the ACA+ALMA only runs show that $\beta$ was unconstrained (see the second panel of Fig.~\ref{fig:corner_highlights}).
        
        % All in all, by adding ACT observations to ACA+ALMA, we directly modeled the pressure distribution on all relevant scales, from the core to half the virial radius in the highest redshift galaxy cluster obtained from the most recent SZ-catalogs.

\section{Two component ICM modeling}\label{sec:twocomp}

    As the merger rate increases with redshift \citep{Fakhouri2010} one might expect a more complex morphology of the pressure distribution in the ICM of XLSSC~122 than a single elliptical component. While Figure~\ref{fig:best_image} indicates that a single component adequately describes the bulk of the ICM, there are still residual features that show negative deviations at $\sim -3\sigma$. Previous studies have demonstrated that on the occasion of a cluster fly-by or when the system is in a premerger or accretion phase, the pressure distribution can exhibit multiple peaks (see, for instance, \citealt{DiMascolo2021}). Hence, these tentative negative features could be attributed to the presence of an infalling group or other kinds of complex morphologies, like filaments. 
    % However, adding multiple SZ components significantly increases the volume of the model's parameter space. To explore this new parameter space, we simplify the model by freezing the slope parameters and eccentricity when modeling the SZ effect via the A10 implementation through ALMA+ACA observations only. To let each run converge to a single solution, we used an ordered prior on the declination for the centroid of the two SZ components, which enforces that one model component has a declination that is always higher than the other. 

    % With the two component modeling, we found a difference in the posterior distribution between profiles that had a steeper inner slope (namely, the A10-CC, A10-UPP, and M14-CC ones) and those with other classifications as the position of the centroid of the posterior distributions modeled with a steeper inner slope showed a higher sampled probability tail ($>1\sigma$) along the southeastern direction peaking at the location of the $-3\sigma$ blob seen in the residuals shown in Figure \ref{fig:best_image}.

    \subsection{ ``Cleaned'' image reconstruction}
        To better highlight possible asymmetric surface brightness distributions,
        % physical insight into the asymmetric posterior distribution,
        we made a ``cleaned'' image reconstruction of the previously shown dirty images of XLSSC~122. The use of the clean algorithm has been the de facto standard in radio astronomy for half a century \citep[e.g.,][]{Hogbom1974}. However, this routine assumes that the emission distribution is well described by an arbitrary set of point-like or multiscale Gaussian sources. Figure~\ref{fig:uv-radial} directly shows that this assumption is invalid in our case, as our emission distribution is unevenly spread along the baselines and mainly concentrated at the smaller \textit{uv}-distances (i.e., large angular scales). Therefore, we constructed our own deconvolution algorithm. Our routine is more analogous to the image reconstruction techniques used in optical interferometry, in which prior information about the source is exploited to deconvolve the dirty beam pattern from the true sky. % \citep{...}. 

        Our routine works as follows: If we assume the signal is well-described by a gNFW profile (instead of a combination of point sources), we can use our forward-modeling technique to find the best-described gNFW profile (think of this as a minor cycle of the deconvolution algorithm) and to subtract this from the visibilities (thus performing only one major cycle). Hence, we do not have an iterative approach but a Bayesian one in which we are guided by the evidence to find the most likely model rather than cleaning to an arbitrary threshold. By subtracting the model from the visibilities, the resulting residual image thus becomes freed of the dirty beam patterns originating from the convolution between the SZ effect and the incomplete \textit{uv}-coverage. Then, similar to clean, we add the imaged residuals (shown on the right panel of Fig.~\ref{fig:best_image}) to the likelihood-weighted model, which is smoothed with the synthesized beam ($5\arcsec$) and attenuated by the primary beam, to create the image reconstruction. Hence, we employ the results of our forward-modeling routine presented above to get a clean-like reconstruction of XLSSC~122. 
        
\begin{figure}[t]
        \centering
        \includegraphics[width = \hsize]{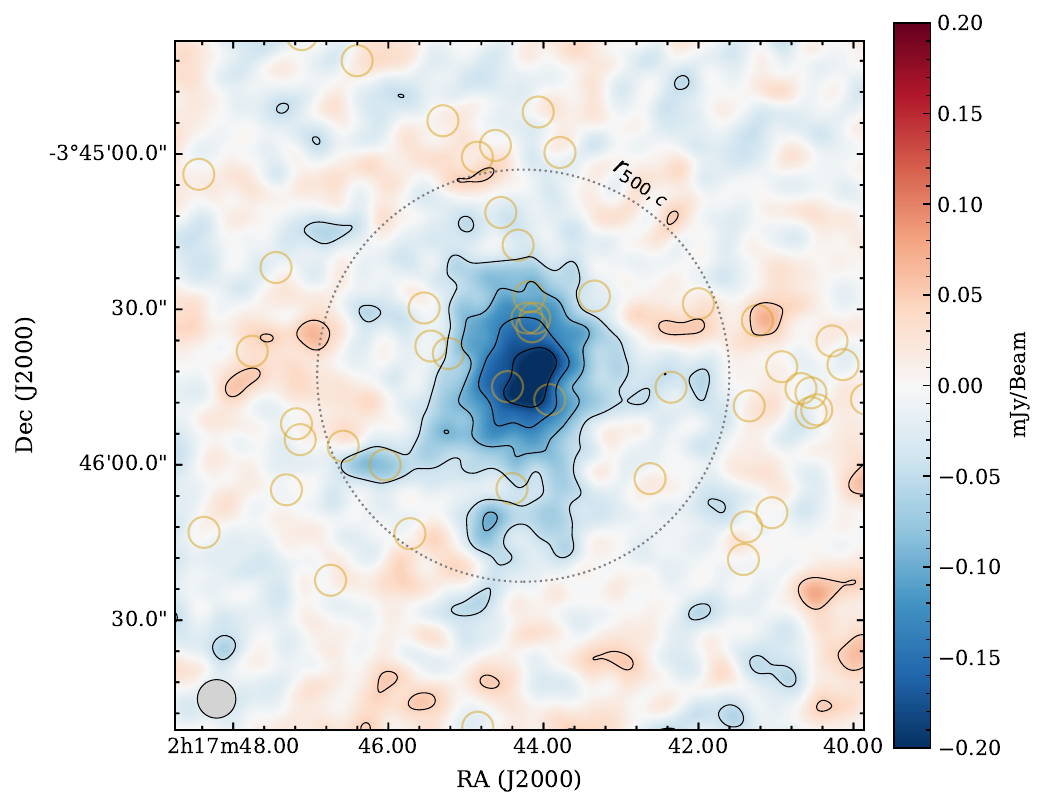}
        \caption{``Cleaned'' ACA+ALMA image of XLSSC~122. Here we combined the likelihood-weighted reconstructed model of a single elliptical gNFW profile (see row five of Table~\ref{table:sigle}), smoothed with the synthesized beam, together with the imaged residuals, computed in the \textit{uv}-plane. Hence, we corrected for the dirty beam patterns visible in Figure~\ref{fig:best_image}. Contours are drawn from [$-10$, $-8$, $-6$, $-4$, $-2$, $2$, $4$]$-\sigma$ estimated on the residual map shown in Figure \ref{fig:best_image}. We overlay the location of the cluster members and indicate $\rm r_{\rm 500,c}$ centered on the peak of the SZ flux. We observe asymmetric features in the south, potentially indicating a morphological disturbance to the cluster.}
        \label{fig:cleaned_image}
    \end{figure}

        The deconvolved ACA+ALMA map of XLSSC~122 is shown in Figure~\ref{fig:cleaned_image}. This image clearly shows two filamentary-like structures in the south which are co-spatial with the negative deviations at $\sim-3\sigma$ shown in the residual map of Figure~\ref{fig:best_image}.
        % , with the one in the southeast being co-spatial with the asymmetric tail in the posterior distribution of the two-component modeling run with the A10-model implementation. This indicates that the forward-modeling routine is finding higher probability solutions along this filament. However, the Bayesian evidence found for the two-component spherically symmetric A10-runs is in the same order ($\Delta |\ln \mathcal{Z}| = 61.7-62.5$) as the single elliptical ones.

    \begin{figure*}[t]
        \centering
        \includegraphics[width = 0.97\textwidth]{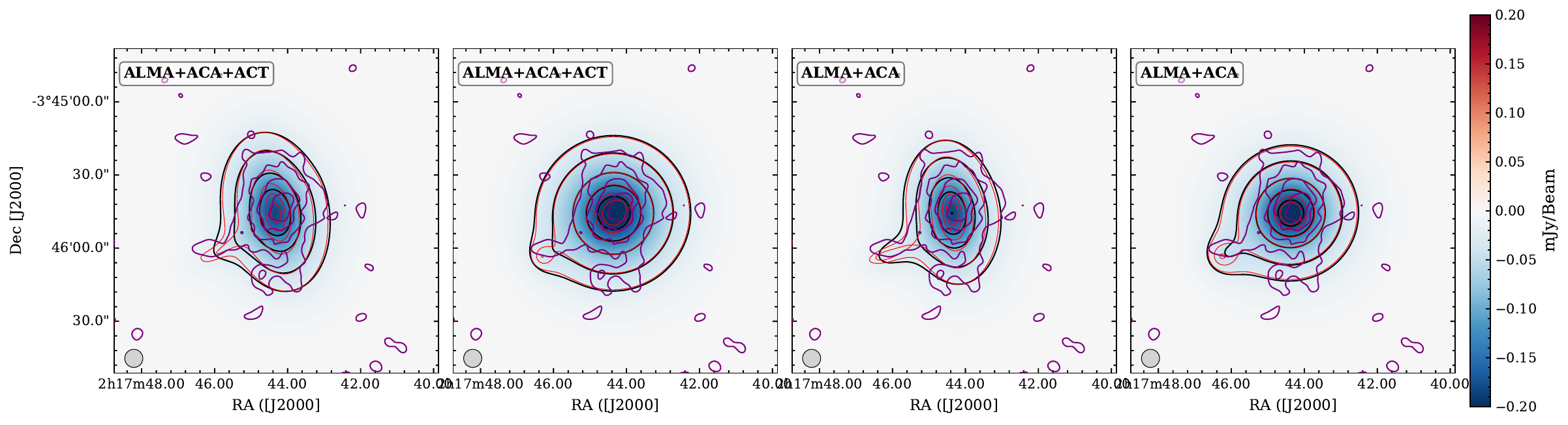}
        \caption{Results on the two-component modeling via the theoretical formalism for both the elliptical (columns one and three) and spherically symmetric implementations (columns two and four). Here, the shape parameters of the gNFW are frozen to the A10-UPP values. The first two columns are modeled to ALMA+ACA+ACT observations and the last two only to ALMA+ACA. In red, we show the unsmoothed model reconstruction. The color map and black contours are the same reconstruction but smoothed with $5\arcsec$ Gaussian taper to match the observations. The contours are drawn at the same levels in Figure~\ref{fig:cleaned_image}. In purple, we show the contours of Figure~\ref{fig:cleaned_image}. This figure indicates that the two-elliptical SZ components better resembles the surface brightness distribution of XLSSC~122 than the spherical one.}
        \label{fig:2comp_modeling}
    \end{figure*}

    \subsection{Modeling asymmetric pressure distributions}
    \label{sec:asymmetric_distributions}

        After obtaining the cleaned interferometric image, we ran a two-component model to confirm via \textit{uv}-based modeling that these filamentary-like structures are not an imaging artifact. The empirical formalism is unsuitable for modeling faint elongated filamentary-like structures due to the constraint between the specific radius and the integrated flux. In contrast, the theoretical formalism decouples the specific radius $r_{\rm s}$ from the amplitude $P_0$, granting greater modeling flexibility for brighter but thinner surface brightness distributions. Hence, we ran the two-component models only with the theoretical formalism.
        
        We consider both a spherical symmetric profile and one in which we freed the eccentricity. The shape parameters of the gNFW are frozen to the A10-UPP values, similar to what is done in Section~\ref{sec:gnfw_single}. To let each run converge to a single solution, we used an ordered prior on the declinations of the centroids of the two SZ components which enforces that one model component has a declination that is always higher than the other. Similar to the single-component runs, we model the two-component runs twice: first is the likelihood computed with only the ACA+ALMA observations, and then with ACT+ACA+ALMA. Thus, in total, we perform four two-component model runs. The resulting four models are shown in Figure~\ref{fig:2comp_modeling}.
        
        Regarding the ACA+ALMA-only modeling, both the spherically symmetric and elliptical models show in the likelihood-weighted image an extended feature along the southeastern filament (see the right two panels of Figure~\ref{fig:2comp_modeling}). For the spherically symmetric model, we observe an improvement of the Bayesian evidence of |$\Delta\ln{\mathcal{Z}}| = 62.3$ relative to the symmetrical one-component fit. For the elliptical one, we find $|\Delta\ln{\mathcal{Z}}| = 68.6$. The latter translates to a Bayesian evidence difference of 6.6 with respect to the best single elliptical gNFW model, indicating a tentative $3.6\sigma$ detection of the second component (see Eq.~\ref{eq:bayesian}). A $\Delta\ln\mathcal{Z} = 6.6$ is considered as decisive evidence \citep{Jeffreys1961}. 

        The likelihood-weighted model reconstructions of the ACT+ACA+ALMA runs are shown in the first two panels of Figure \ref{fig:2comp_modeling}. When modeling with the additional ACT observations, only the elliptical implementation clearly shows the asymmetric feature along the southeastern direction. From the Bayesian evidence, we find $\Delta|\ln{\mathcal{Z}}| = 116.8$ and $\Delta|\ln{\mathcal{Z}}| = 121.8$, which corresponds to a difference of $-0.4$ and $3.0$ (with the latter equivalent to a tentative $2.4\sigma$ preference) when compared with their respective single component gNFW model for the spherical and elliptical implementations. This implies that the observations require some degree of elongation in both the north-south and the cross-diagonal orientation. The absence of a secondary component in the spherically symmetric two-component ALMA+ACA+ACT run, as opposed to the ALMA+ACA run, and the lower significance of the ALMA+ACA+ACT run with respect to the ALMA+ACA run can be attributed to the flux constraint imposed by the ACT observations on the overall system. The ALMA+ACA observations lack short-baseline information, providing greater maneuverability for the model as demonstrated in Figure~\ref{fig:2comp_modeling} where the integrated flux in the spherical symmetric ALMA+ACA run is larger when compared to the other. 

        Regarding the eccentricity, the modeling done on both the ACA+ALMA and ACA+ALMA+ACT observations imply an eccentricity for the smaller component of $\sim0.8$. This is extreme; however, the cause is most likely because of the model choice. Currently, there are no accurate models to describe these filamentary-like structures which were computationally feasible. However, to overcome a possible mismatch between the model implementation (a very eccentric gNFW profile) and the tentative filamentary-like structure, we weighted each sample of the nested sampling routine with its likelihood and computed the weighted average, which is the same as marginalizing over the posterior distribution. Then, we take into account all the complex degeneracies seen in the posterior distribution.
        Figure~\ref{fig:2comp_modeling} shows each of the marginalized two-component gNFW models.
        By marginalizing over the posterior distribution, we get a smoother model which closely resembles the cleaned image, as indicated in Figure \ref{fig:2comp_modeling} where we plot the contours from the cleaned image reconstruction (see Fig.~\ref{fig:cleaned_image}) on top of the models. We find that the southern filament is part of the bulk of the ICM, while the southeastern elongated structure is best described by the secondary, smaller component. The integrated flux of the secondary component is roughly twice as faint as the bulk of the ICM, and the projected centroids are separated by $\approx33\arcsec$. 

\section{Mass estimates}\label{sec:masses}

\begin{figure}[t]
            \centering
            \includegraphics[width = \hsize]{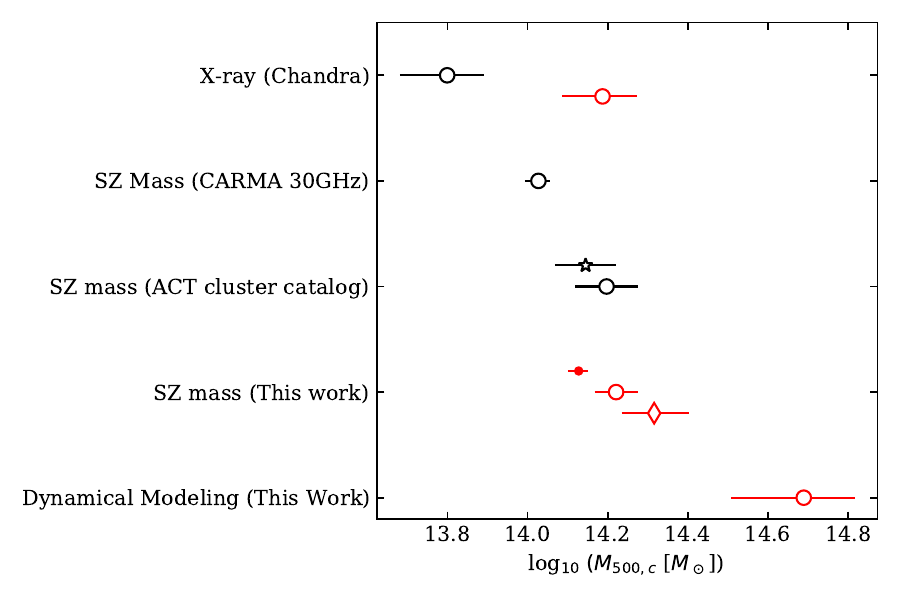}
            \caption{Overview of halo masses ($M_{500,\mathrm{c}}$) of XLSSC~122. In red, we show the halo masses derived in this work. Regarding SZ measurements, the red open circle corresponds to the likelihood-weighted average of all masses derived from the empirical formalism (Table~\ref{table:sigle_ACA_ALMA_ACT}, lower table). The dot represents the mass estimates from the spherically symmetric model and the diamond that of elliptical ones. For the halo masses reported in the ACT cluster catalog, the star shows the calibrated mass estimate, and the circle the SZ mass obtained by matched filtering the frequency maps. The reported errors correspond to the $16^{\rm th}-84^{\rm th}$ quantiles. This figure shows the discrepancies between the derived $M_{500,\mathrm{c}}$ indicating that with forming clusters, one cannot reliably use one tracer or constant gas fraction to estimate the true halo mass.  }
            \label{fig:mass_compare}
        \end{figure}

        Figure \ref{fig:mass_compare} shows an overview of halo masses ($M_{500,\mathrm{c}}$) of XLSSC~122. Our SZ mass is calculated by taking the evidence-weighted average of the mass estimates from the empirical formalism runs utilizing ALMA+ACA+ACT observations as shown in Table~\ref{table:sigle_ACA_ALMA_ACT}, and is equal to $M_{500,\mathrm{c}} = 1.66^{+0.23}_{-0.20} \times 10^{14}~\rm M_\odot$. Here, we make use of local scaling relations from \citet{Arnaud2010} to convert $Y_{\rm 500,c}$ to an SZ-derived halo mass, as robust scalings derived from high$-z$ clusters do not exist.
        
        The halo masses of XLSSC~122 from \citet{Mantz2018} are derived from X-ray imaging (\textit{Chandra}) and CARMA measurements. First, the X-ray derived halo mass reported by \citet{Mantz2018} was obtained by integrating their radial density profile while using an X-ray spectroscopic temperature of $kT = 5.0$~keV and adopting a gas mass fraction of $f_{\mathrm{gas}}(r_{500,\mathrm{c}}) = 0.125$. We made an independent estimate of the halo mass using these same data based on the $M_{\rm 500,c}-T_{\rm X}$ relationship of \citet{2009ApJ...692.1033V}. Adopting the temperate of $T_{\rm X} = 5.0 \pm 0.7$~keV, we obtain $\log_{10}\left(M_{\rm 500,c} \left[~M_\odot\right] \right)= 14.19\substack{+0.09 \\ -0.10}$ (also shown, in red, in Fig.~\ref{fig:mass_compare}). The errors are propagated from the temperature information and do not include any systematics. We note that this mass estimate is much more in line with the SZ-derived halo mass than converting the X-ray-derived gas mass to a halo mass as done by \citet{Mantz2018}.

        Further, we converted the $Y_{500,\mathrm{c}}$ estimate from \citet{Mantz2018} -- which was obtained by fitting an A10-UPP model to their CARMA observations -- using the same methodology employed for our mass estimates; that result is shown as ``CARMA 30 GHz'' in Figure~\ref{fig:mass_compare}. We also quote the derived cluster mass of XLSSC~122 as reported in the ACT cluster catalog \citep{Hilton2021}. We show both the mass derived from the matched filter and one with an additional correction term \citep[see][for details]{Hilton2021}. Finally, we add estimates on the halo mass derived via the dispersion-velocity measurements of the cluster members XLSSC~122 dynamical mass.

        To derive the dynamical mass, we adopt the methodology described in \citet{2022A&A...659A.126A}. We used the cluster members in \citet{Willis2020} that were labeled as the ``Gold'' standard to derive the velocity dispersion. In total, we use the spectroscopic redshifts of 32 cluster members that fall within $r_{200,\mathrm{c}}$. We found a $\sigma_{200,\mathrm{c}} = 1014~\pm169$ km s$^{-1}$ by converting the redshift differences between cluster members and the median redshift to velocity offsets and estimating the biweight scale \citep{1990AJ....100...32B} and standard error over these velocities. Then, we used the scaling of \citet{2013MNRAS.430.2638M}:

        \begin{equation}
            \frac{M_{200,\mathrm{c}}^{\mathrm{dyn}}}{1\times10^{15}~\mathrm{M_\odot}} = \left(\frac{\sigma_{200,\mathrm{c}}}{A}\right)^{\frac{1}{\alpha}}, 
        \end{equation}

        \noindent with $A = 1177.0$ km s$^{-1}$ and $\alpha = 0.364$ to convert the velocity dispersion to a dynamical mass estimate. To convert the halo mass to the $M_{500,\mathrm{c}}$ definition we use the concentration parameter definition of \citet{2019ApJ...871..168D} and the $M_{200,\mathrm{c}}^{\mathrm{dyn}}$ to $M_{500,\mathrm{c}}^{\mathrm{dyn}}$ conversion as implemented by the \texttt{colossus} package \citep{2018ApJS..239...35D}. To estimate the uncertainty of the dynamical mass, we adopt the procedure of \citet{2020A&A...641A..41F}:
                
        \begin{equation}
            \Delta M_{200,\mathrm{c}}^{\mathrm{dyn}} = M_{200,\mathrm{c}}^{\mathrm{dyn}} \sqrt{\frac{\epsilon}{4(N_\mathrm{gal}-1)^\beta}},
        \end{equation}

        \noindent with $\epsilon = 16.2$ and $\beta = 1.13$. Here, we only propagated the statistical error on the velocity dispersion to the dynamical mass uncertainty as the redshift uncertainty was not given by \citealt{Willis2020}. All peculiar velocities were estimated with respect to the BCG, which was selected as the galaxy with the lowest magnitude. Following this procedure we find $\log_{10} M_{500,\mathrm{c}}^{\mathrm{dyn}} = 14.69^{+0.13}_{-0.18}~\rm M_\odot$. 

        We find lower SZ-derived mass estimates than what is expected from the velocity-dispersion measurement ($\sim\times2.5$). This finding is in line with that of other high-redshift SZ detections \citep{DiMascolo2023,2023MNRAS.522.4301A}. It suggests that the hot component of the ICM is still assembling and is actually part of several interacting substructures of lower-density gas. This hypothesis agrees with the lower X-ray derived mass estimate: because the X-ray emissivity is proportional to $\epsilon\propto n_\mathrm{e}^2$, it becomes decreasingly sensitive to lower dense regions, while the SZ effect is linearly proportional to the electron density, $n_\mathrm{e}$, and thus more sensitive to the hot gas component of the ICM in forming clusters. Hence, the lower halo mass estimates derived from X-ray observations are consistent with the picture that XLSSC~122 is still actively forming and largely composed of lower-density gas. If this is the case, the gas fraction of $f_\mathrm{gas} = 0.125$ used in \citet{Mantz2018} is most likely underestimating the true mass. 
        
    \section{Discussion} \label{sec:discussion}
    
        \subsection{Morphological implications}

        \begin{figure*}[t]
            \sidecaption
            % \centering
            % \vspace{1cm}
            \includegraphics[width=0.7\textwidth, trim={0 2mm 0 0},clip]{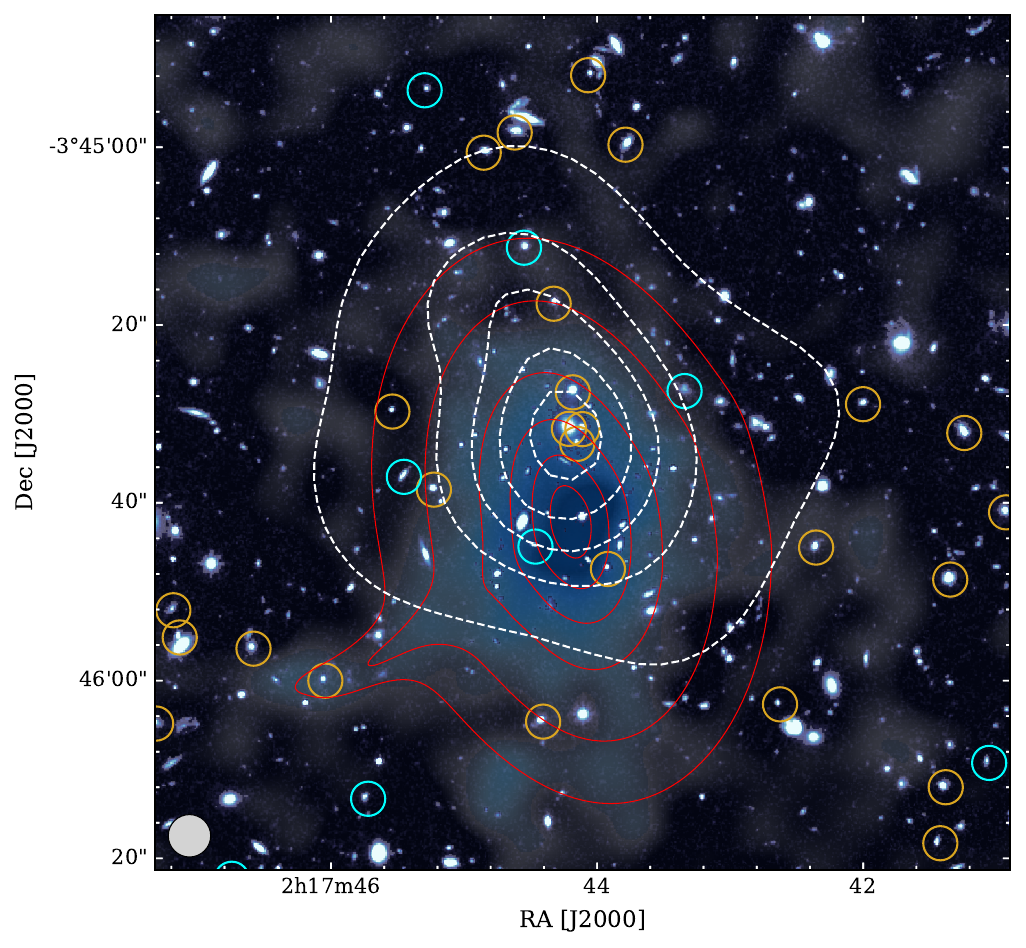}
            % \vspace{-55mm}
            \caption{Complete multiwavelength view of XLSSC~122. We show adaptively smoothed \textit{XMM-Newton} contours (white) overplotted on the HST F140W background. The cluster members are highlighted in gold and cyan, with the latter corresponding to star-forming and dusty galaxies based on the SED-fitting done by \citet{2022MNRAS.515.2529T}. In between the cluster members, we visualize the SZ flux ($<-2\sigma$) as seen in the cleaned ALMA+ACA observations in blue. The beam size of the ALMA+ACA image (tapered to $5\arcsec$) is shown in the bottom left. The red contours represent our most likely model reconstruction obtained by forward modeling two gNFW components to the ALMA+ACA+ACT observations. The figure indicates a clear morphological difference between the equation of state parameters traced by X-ray and the SZ effect.}
            % \vspace{55mm}
            \label{fig:awesome}
        \end{figure*}
        
        The main question of this paper is what phase of cluster assembly XLSSC~122 is in and, correspondingly, how disturbed it is. We provide an answer by combining high-resolution ALMA+ACA Band 3 observations with ACT data. By jointly modeling ALMA+ACA+ACT, we find that XLSSC~122 can be classified as a non cool-core when compared to local observations or as relaxed when compared to distant clusters. However, the difference between these two profile classifications can be attributed to the increasing merger rate at higher redshifts \citep{Fakhouri2010}, which logically explains the flattening of the profile. Even though the terminology used in \citet{McDonald2014} conveys the idea that XLSSC~122 can be interpreted as a ``cool-core'' cluster based on its pressure profile, it is actually undergoing some degree of morphological disturbance.
        
        This morphological disturbance is consistent with the modeled eccentricity of $e=0.46^{+0.12}_{-0.16}$. Even though the eccentricity strongly depends on the projection of the merger/post-merger on the sky \citep[see, e.g.,][]{2018MNRAS.477..139C} that can make an elliptical structure look spherically symmetric, higher ellipticities must be caused by a morphological disturbance which is unresolved in the observations. 

        On slightly larger scales, we do resolve asymmetric features in the south of the cluster. By modeling these filamentary-like structures with a highly elliptical gNFW model, we tentatively confirm---at $3.6\sigma$ for ALMA+ACA and at $2.4\sigma$ when ACT is included (see Section~\ref{sec:asymmetric_distributions})---the presence of a second component in the pressure distribution. We are aware that this specific model is not ideal for modeling filamentary-like structures, but it is sufficient to probe the significance of this feature in the native, \textit{uv}-space of ACA and ALMA, rather than trying to infer a physical interpretation of the $\sim-3\sigma$ feature in the residual map. Future work should explore a wider variety of models that can capture the complexity of these asymmetric features, which will most likely increase the detection inference of the second component. However, the tentative presence of this second component, which could be an infalling group, a filament, or a projection effect of an asymmetric assembling cluster, does strengthen the idea that this cluster is actively assembling and does not follow the definition of a relaxed cool-core-like structure.

        By overlaying the cleaned SZ flux with X-ray emission and cluster member distribution, as shown in Figure \ref{fig:awesome}, we notice two things: first, there is a large excess of gas in the south of the cluster where no X-ray emission is detected. This is in line with the difference in the reported mass estimates derived from X-ray and the SZ effect as mentioned in Section \ref{sec:masses}, again strengthening the idea that the hot ICM in XLSSC~122 is composed of low-density gas and is actively growing. Second, the peak of the pressure distribution, as mapped by the SZ signal, is offset from the BCG and the peak of the X-ray emission by $10\arcsec\pm 1\arcsec$ ($80\pm7$ kpc) when projected on the sky mainly along declination axis.\footnote{We note that the ALMA measurements still contradict the CARMA results on the centroid position by $20\arcsec\pm1\arcsec$ (corresponding to a $3\sigma$ difference based on their derived uncertainty of the centroid, \citealt{Mantz2018})}. The following section provides details on the possible implications.

        \subsection{Thermodynamical implications}\label{sec:south}
        
        To understand the offset between the SZ peak and the BCG, which is co-spatial with the X-ray imaging, as well as the excess of SZ flux in the southern part of the cluster, we can compare XLSSC~122 to the well-studied local cluster ($z=0.451$) RX J1347.5-1145. This cluster is one of the brightest X-ray emitting galaxy clusters which has been studied extensively over the last decade in the mm-wave regime (see e.g.\ \citealt{2001PASJ...53...57K}, \citealt{Kitayama2016}, \citealt{2018ApJ...866...48U}, and \citealt{DiMascolo2019a}). RX J1347.5-1145 hosts an excess of SZ flux in the southeast of the X-ray peak that is most likely a strong, shock-induced pressure perturbation caused by a major merger event. Similar to  XLSSC~122, this excess is found at an offset of $27\arcsec$ ($\sim109h^{-1}$ kpc) from the X-ray peak and is faint in the X-ray but bright in the SZ signal.

\begin{figure}[t]
            \centering
            \includegraphics[width=0.99\hsize]{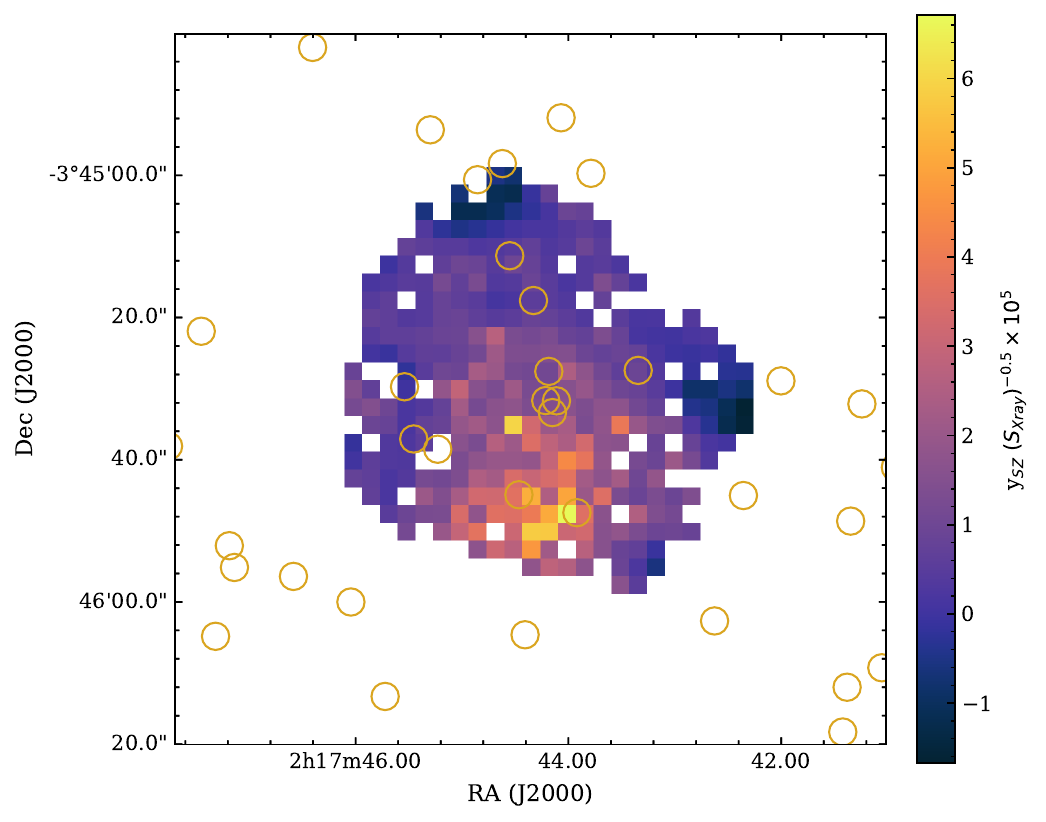}
            \caption{Pseudo-k$_\mathrm{B}$T$_\mathrm{e}$ ($y/\sqrt{S_\mathrm{X}}$) map of XLSSC~122. It takes the ratio of the cleaned SZ map (Fig.~\ref{fig:cleaned_image}) and the X-ray surface brightness (2$^{\textnormal{nd}}$ panel of Fig.~\ref{fig:multiwavelength}) within the $2\sigma$ adaptively smoothed X-ray contours (also shown in Fig.~\ref{fig:multiwavelength} and Fig.~\ref{fig:awesome}). The units are linearly scaled to an arbitrary value to show the relative pseudo temperature change in the south of XLSSC~122.}
            \label{fig:ratio}        
        \end{figure}

        Since X-ray spectroscopy provides a temperature constraint that is essentially emission weighted, and the bulk of the X-ray emission is seen in the northern portion of the cluster, the global temperature constraint of $T_{\textnormal{spec}} = 5$ keV \citep{Mantz2018} may not be valid in the southern portion of the cluster where the SZ signal peaks. 
        Generally, merger events can temporarily boost the SZ flux on time scales much smaller than the virialization time of a massive cluster \citep{wik2008}. Gas in an infalling substructure can be stripped by ram pressure of the main bulk of the ICM and create pressure perturbations induced by shock waves which boost the $y$-value locally. To investigate this further, we estimated temperature differences in XLSSC~122 by taking the ratio of the SZ flux over the square root of the X-ray surface brightness. Since the SZ effect traces the integrated pressure along the line of sight, and the X-ray emission traces roughly the density squared, combining them in this way results in a quantity proportional to the temperature of the ICM via the ideal gas law (under the assumption of constant temperature along the line of sight). One would expect a constant ratio if the gas is isothermal and both the X-ray and SZ observations have a similar point spread function (PSF). Figure~\ref{fig:ratio} shows this ratio scaled to arbitrary units and indicates that XLSSC~122 is far from isothermal, with a pseudo temperature increase of $\sim3\times$ in the south, indicative of a disturbance. In Figure~\ref{fig:ratio}, both the \textit{XMM-Newton} observations and the ALMA+ACA image reconstruction have a resolution of $\approx5\arcsec$.

        The pseudo temperature increase in the south is co-spatial with the tentative second SZ component and suggests gas stripping of an infalling subcluster which shock-heated the gas. 
        % This scenario is in line with the findings of a ``second'' BCG located in the south of the cluster. Since this second-most massive quenched galaxy is located at the south of the cluster at RA $= 2\fh17\fm44\fs4720$, dec = $-3\fdg46\farcm04\farcs800$, and $\Delta z = 0.007$ from the BCG \citep{2022MNRAS.515.2529T}, we speculate that it belongs to the infalling group.
        This putative merger of an infalling group in the large-scale filamentary structure with the bulk of the already-formed ICM could temporarily boost the $Y$-value causing it to exceed the ACT detection threshold and thus be included in the catalog of \citep{Hilton2021}. 
        However, other mechanisms may also contribute to explaining why XLSSC~122 is the only cluster detected by ACT during this epoch, despite having a halo mass well below the limit set by the hierarchical growth model of dark matter halos. For instance, bright radio galaxies are often located at the cores of (proto-)clusters \citep[see, e.g.,][]{DiMascolo2023} and reduce the amplitude of the SZ flux by infilling it with a positive signal, but XLSSC~122 lacks a radio-loud AGN in its BCG and thus its SZ signal is uncontaminated. Another possible explanation is the lack of optical confirmations of high$-z$ cluster candidates ($z>1.3$), which would exclude them from the catalog of \citep{Hilton2021}.

        In summary, our findings and interpretation on the forming ICM are: 
        (1) we have tentative evidence for a second pressure component with a flux ratio of $\sim1{:}2$; 
        (2) there is an offset between the pressure peak and the X-ray peak, which causes the enhancement of the pseudo temperature in the south; 
        (3) the fact that XLSSC~122 is the only optically confirmed cluster of its epoch detected by ACT could be because of a merger event in which the substructure is going through its first core passage, temporarily boosting the $y$-value \citep{wik2008}; 
        % (4) the second-most massive quenched galaxy is located at the south of the cluster which hints at the presence of a ``second'' BCG falling in from the cosmic web; 
        and (4) the mismatch between X-ray and SZ-derived halo masses suggests that XLSSC~122 is undergoing a major merger that is heating the ICM, as opposed to heating coming from constant, small-scale accretion from the cosmic web.
       
    \subsection{Time scales}
        We estimate time scales to gauge the evolutionary stage of XLSSC 122. At any redshift, the masses of the largest gravitationally bound objects are sensitive to the underlying cosmology. The earliest objects of a given kind form at the rare, high peaks of the density distribution of the Universe \citep{1974ApJ...187..425P}. At a redshift of $z\sim2$, the most massive and newly formed structures are (proto)clusters. From cosmology, we know that the density power spectra peaks at roughly $8 h^{-1}$ Mpc. As a sphere with a radius of $8 h^{-1}$ Mpc contains about the right amount of material to form a cluster \citep{Peebles1980, 2014arXiv1401.1389L}, we can find how far XLSSC~122 is in its assembly phase by estimating how large a sphere has already collapsed into the structure we observe.

        If we assume the tentative detection of the colliding substructure in the south of the cluster as true, then by simply assuming a peculiar velocity of 1000 km~s$^{-1}$ $\simeq$ 1 Mpc~Gyr$^{-1}$, we can estimate via standard Newtonian physics how large a sphere has already collapsed into what we know as XLSSC~122. If we treat the structures as point-like objects in which the infalling substructure accelerates constantly from $v_\textrm{pec} = 0$ km~s$^{-1}$ to $v_\textrm{pec} = 1\times 10^3$ km~s$^{-1}$ within the expansion time of the Universe (which is $t = 3.3$ Gyr at $z=1.98$), we find that the distance the substructure has traveled is

        \begin{equation*}
            \Delta x = \frac{\Delta v \Delta t}{2} \simeq 1.8~\textnormal{Mpc},
        \end{equation*}

        \noindent which falls well within the $\sigma_8$-radius. Of course, observations cannot be fast-forwarded in time, and the only true indicator of whether XLSSC~122 can grow to become a local Coma-like cluster is its large scale surrounding \citep{Remus2023}. However, even though the above estimate is simplistic, it sketches the idea that XLSSC~122 still has the potential to grow.
        
        Furthermore, we can estimate the virialization time and compare it to the lifetime of the Universe. The virialization time is roughly equal to three to ten times the sound crossing speed \citep{wik2008}. Using the global ICM temperature form \citet{Mantz2018} we find the time to virialize

        \begin{equation*}
            t_\mathrm{vir} = \frac{3R}{c_\mathrm{s}} = 3R\left(\sqrt{\gamma \frac{k_\mathrm{B}T}{m}}\right)^{-1} \simeq \frac{3R}{0.0038 c} \simeq 0.5~\textnormal{Gyr},
        \end{equation*}

        \noindent for $c_\mathrm{s}$ the sound speed, $c$ the speed of light, $\gamma = 5/3$, and $m = \mu m_\mathrm{p}$, which is the proton mass multiplied by the mean molecular weight (or mass per particle). Thus a time scale fitting comfortably within the age of the Universe at that epoch provides sufficient time for the gas to convert its gravitational potential energy into thermal energy, which in turn enables one to detect it through observations of the SZ effect.

        Even though XLSSC~122 is the most distant cluster detected in the ACT-cluster catalog, independent follow-ups of overdensities of galaxies around radio-bright AGN have found protoclusters at more distant redshifts. As an example, \cite{DiMascolo2023} reports the first robust detection of the forming hot ICM component in a (proto-)cluster at $z>2$, also using ALMA+ACA observations of the SZ effect. In that case, the target was the Spiderweb proto-cluster, which at $z=2.156$ lies 300 Myr further back in cosmic time than XLSSC~122. Through modeling an A10-MD profile to the observations, \citet{DiMascolo2023} found a $Y(<r_{500,\mathrm{c}}) = 0.76^{+0.19}_{-0.17}~\times~10^{-6}~\textnormal{Mpc}^2$. By fitting the same model to XLSSC~122, we find $Y(<r_{500,\mathrm{c}}) = 2.0^{+0.6}_{-0.4} \times 10^{-5}~\textnormal{Mpc}^2$, approximately 20$\times$ the intrinsic $Y(<r_{500,\mathrm{c}})$ of the Spiderweb. Furthermore, XLSSC 122 exhibits a red sequence \citep{Willis2020, Noordeh2021}, while the Spiderweb proto-cluster is composed more of star-forming cluster members.
        
        From the foregoing time scale estimates and by comparing with the Spiderweb protocluster, we propose that if the Spiderweb can be said to be in its ``early childhood'' phase, XLSSC~122 would be an ``adolescent'' cluster. It is still assembling and constitutes a bona fide yet immature galaxy cluster.
        
\section{Conclusions}\label{sec:conclusion}

    In this work, we add high-resolution ($\approx 5\arcsec$) ALMA (12m-array) and ACA (7m-array) Band 3 observations to augment an extensive collection of auxiliary data on XLSSC~122, the most distant cluster detected in recent cluster SZ catalogs. Through forward modeling analytical prescriptions of the pressure distribution to the interferometric ALMA+ACA observations jointly with those made by the single dish telescope ACT, we model the pressure distribution from the core ($\approx5\arcsec$, $\sim 40$ kpc) to roughly half the virial radius. The results obtained from our forward modeling analysis lead us to the following conclusions:
   
    \begin{enumerate}
        \item We detect the SZ effect with a significance of $11\sigma$ in the ALMA+ACA data alone which increases to $15\sigma$ when ACT observations are included in the forward modeling routine. The significance is determined through the Bayesian evidence, as presented in Tables~\ref{table:sigle}~\&~\ref{table:sigle_ACA_ALMA_ACT}. 
        Notably, in comparison to prior follow-up observations with CARMA, measurements using ALMA+ACA+ACT have higher resolution, sensitivity, and dynamic range, which both allow for higher fidelity imaging and improve the mitigation of contamination by compact sources. The result is that we find better agreement between the SZ decrement and the X-ray emission seen in archival {\it XMM-Newton} and {\it Chandra} observations.
        \item Based on its radial pressure distribution, XLSSC~122 is classified as a noncool-core when compared with local observations \citep[see][]{Arnaud2010}. In contrast, when compared to profiles of more distant clusters of galaxies \citep[][]{McDonald2014}, XLSSC~122 exhibits a relatively relaxed state. However, via the Bayes factor, we cannot distinguish between the two classifications, but we can distinguish them from the other pressure profile templates from \citet{Arnaud2010} and \citet{McDonald2014} with a significance of $\Delta\ln\mathcal{Z} \geq 4.5$ ($\geq 3\sigma$, see Table~\ref{table:sigle_ACA_ALMA_ACT}).
        \item XLSSC~122 exhibits an eccentric structure, with $e=0.46^{+0.12}_{-0.16}$, also indicating a morphologically disturbed nature of the cluster. Furthermore, our analysis leads to an improved precision of the SZ mass estimate to $M_{500,\mathrm{c}} = 1.66^{+0.23}_{-0.20} \times 10^{14}~\rm M_\odot$, though we note the overall accuracy could still be affected by the hydrostatic mass biases common to SZ mass estimates.
        \item By reconstructing the interferometric image with the marginalized model reconstruction analog to the clean algorithm, we found an excess of SZ flux in the south with respect to the BCG and the X-ray surface brightness. Then, through modeling the SZ surface brightness with two components, we tentatively confirm the presence of a second source or filamentary-like structure to the southeast with $\Delta\ln{\mathcal{Z} = 6.6}$  ($\sigma_{\rm eff} = 3.6$) when modeling the ALMA+ACA observations alone and $\Delta\ln{\mathcal{Z} = 3.0}$ ($\sigma_{\rm eff} = 2.4$) when including ACT observations. We speculate that this second component could boost the Compton $Y$ value locally as the gas is heated. As the cluster is still actively forming, the gas of the hot ICM is relatively low in density; hence the excess of gas in the south is detected with the SZ effect while going unnoticed in the X-ray wavelengths. 
        \item By comparing XLSSC~122 to local observations and even more distant clusters of galaxies, we posit the idea that XLSSC~122 is in its ``adolescent'' phase. Even though detected at $z\sim2$, XLSSC~122 had time to virialize and attract matter over time, forming a bulk of hot but low-density gas in the ICM. Through our multiwavelength approach, we believe that XLSSC~122 is likely undergoing a major merger and that a major mechanism driving the heating of the ICM could be through this collision rather than constant small-scale accretion of matter from the cosmic web. This collision would have boosted the $Y$-value temporarily, causing it to exceed the ACT detection threshold, making it the only cluster around $z\sim2$ that is optically confirmed and detected in the southern hemisphere by ACT.
    \end{enumerate}

    Regarding future work in this field, we anticipate that ALMA imaging and characterization of the forming hot ICM via the SZ effect is only now beginning in earnest with the introduction of ALMA Band 1 \citep[35-50~GHz; see][]{ALMAband12013, Huang2022}. With 13 times the collecting area and a larger field of view, while sampling similar spatial scales as the ACA in Band 3, we can start to resolve the ICM in the most distant clusters of galaxies with only a fraction of the integration time. This is timely as Simons Observatory will come online next calendar year \citep{2019JCAP...02..056A}. This single-dish CMB survey telescope is expected to find more and more high-z clusters which can be followed up to begin building a statistical sample of resolved observations to understand ICM heating, cluster growth, and evolution. Looking to the further future, we can expect significant advances to be provided by major new (sub-)millimeter facilities, such as the 50-meter Atacama Large Aperture Submillimeter Telescope (AtLAST; \citealt{Ramasawmy2022,mroczkowski2023, 2024arXiv240218645M}). AtLAST will feature a large $2^\circ$ FOV with a $10\arcsec$ resolution at 150~GHz, providing a more complete, high spatial dynamic range view of the SZ effect in clusters, bridging the detailed view provided by ALMA and the nearly all-sky view from CMB experiments.

\begin{acknowledgements}
     We thank A. Saro and T. Kitayama for the fruitful conversations and guidance in the course of this work. We also thank A. Mantz and J. Zavala for proposing the archival ALMA/ACA observations used here.
     
    Support for ACT was through the U.S.\ National Science Foundation through awards AST-0408698, AST-0965625, and AST-1440226 for the ACT project, as well as awards PHY-0355328, PHY-0855887 and PHY-1214379. Funding was also provided by Princeton University, the University of Pennsylvania, and a Canada Foundation for Innovation (CFI) award to UBC. ACT operated in the Parque Astron\'omico Atacama in northern Chile under the auspices of the Agencia Nacional de Investigaci\'on y Desarrollo (ANID). The development of multichroic detectors and lenses was supported by NASA grants NNX13AE56G and NNX14AB58G. Detector research at NIST was supported by the NIST Innovations in Measurement Science program. Computing for ACT was performed using the Princeton Research Computing resources at Princeton University, the National Energy Research Scientific Computing Center (NERSC), and the Niagara supercomputer at the SciNet HPC Consortium. SciNet is funded by the CFI under the auspices of Compute Canada, the Government of Ontario, the Ontario Research Fund–Research Excellence, and the University of Toronto. We thank the Republic of Chile for hosting ACT in the northern Atacama, and the local indigenous Licanantay communities whom we follow in observing and learning from the night sky.

     This paper makes use of the following ALMA data: ADS/JAO.ALMA\#2016.1.00698.S and ADS/JAO.ALMA\#2018.1.00478.S. ALMA is a partnership of ESO (representing its member states), NSF (USA) and NINS (Japan), together with NRC (Canada), MOST and ASIAA (Taiwan), and KASI (Republic of Korea), in cooperation with the Republic of Chile. The Joint ALMA Observatory is operated by ESO, AUI/NRAO and NAOJ.

     ADH acknowledges support from the Sutton Family Chair in Science, Christianity and Cultures, from the Faculty of Arts and Science, University of Toronto, and from the Natural Sciences and Engineering Research Council of Canada (NSERC) [RGPIN-2023-05014, DGECR-2023-00180].

     KM acknowledges support from the National Research Foundation of South Africa.

     TM acknowledges support from the AtLAST project, which has received funding from the European Union’s Horizon 2020 research and innovation program under grant agreement No 951815.

     LDM is supported by the ERC-StG ``ClustersXCosmo'' grant agreement 716762 and acknowledge financial contribution from the agreement ASI-INAF n.2017-14-H.0.

     CS acknowledges support from the Agencia Nacional de Investigaci\'on y Desarrollo (ANID) through Basal project FB210003.

     The project leading to this publication has received support from ORP, which is funded by the European Union’s Horizon 2020 research and innovation program under grant agreement No 101004719 [ORP].

\end{acknowledgements}

\bibliographystyle{aa} % style aa.bst
\bibliography{XLSSC} % your references Yourfile.bib

\begin{appendix}

\onecolumn

    \section{Priors}\label{app:cornerplots}
        The posterior distributions for all runs shown in Table~\ref{table:sigle}~\&~\ref{table:sigle_ACA_ALMA_ACT} are shown in the supplementary material. The priors used for these runs are shown here in Table \ref{tab:priors}.

        \begin{table}[h]
            \centering
            \resizebox{\textwidth}{!}
            {%
            \begin{tabular}{lllclllc}
            \toprule
            \toprule
                                   & Parameter              & Prior Type  & Min -- Max                 &                         & Parameter              & Prior Type  & Min -- Max              \\ 
            \midrule
            \textit{Empirical formalism} &                         &             &                            & \textit{Theoretical-formalism} &                        &             &                        \\ 
                                   & RA [deg]                & Uniform     &  34.425921 -- 34.442521    &                         & RA [deg]               & Uniform     &  34.425921 -- 34.442521\\ 
                                   & Dec [deg]               & Uniform     &  -3.768419 -- -3.751819    &                         & Dec [deg]              & Uniform     &  -3.768419 -- -3.751819\\ 
                                   & $M_{500,\mathrm{c}}$ [M$_\odot$] & Log-Uniform &  $10^{13.4} - 10^{14.6}$   &                         & $P_0$ [keV~cm$^{-3}$]     & Uniform     &  0.0001 -- 1.0         \\ 
                                   & eccentricity              & Uniform     &   0.0 -- 1.0               &                         & $r_{\rm s}$ [arcsec]         & Uniform     &  1.1772 -- 120.024     \\ 
                                   & Position angle  [deg]   & Uniform     &   -90.0 -- 90.0            &                         & eccentricity            & Uniform     &  0.0 -- 1.0            \\ 
                                   & $z$                     & Gaussian    &  1.977 -- 1.979            &                         & Position angle [deg]   & Uniform     &  -90.0 90.0            \\ 
                                   & $\alpha_\mathrm{ACA, B3}$      & Gaussian    &  0.95 -- 1.00              &                         & $\alpha$               & Uniform     &  0.5--10.0             \\ 
                                   & $\alpha_\mathrm{ALMA, B3}$     & Gaussian    &  0.95 -- 1.00              &                         & $\beta$                & Uniform     &  0.5--10               \\ 
            \textit{Point Source}  &                         &             &                            &                         & $\gamma$               & Uniform     &  -1.0 -- 5.0           \\ 
                                   & RA  [deg]               & Uniform     & 34.42588105 -- 34.45717415 &                         & $z$                    & Gaussian    &   1.977 -- 1.979       \\ 
                                   & Dec [deg]               & Uniform     & -3.7905406 -- -3.7377484   &                         & $\alpha_\mathrm{ACA, B3}$     & Gaussian    &  0.95 -- 1.00          \\ 
                                   & Amplitude [Jy]          & Uniform     & 0.0 -- 1.0                 &                         & $\alpha_\mathrm{ALMA, B3}$    & Gaussian    &  0.95 -- 1.00          \\ 
                                   & Spectral slope          & Uniform     & -5.0 -- 10   \\
                                   & $\alpha_\mathrm{ACA, B3}$      & Gaussian    & 0.95 -- 1.00 \\
                                   & $\alpha_\mathrm{ALMA, B3}$     & Gaussian    & 0.95 -- 1.05 \\ 
                                   & $\alpha_\mathrm{ALMA, B4}$     & Gaussian    & 0.95 -- 1.05 \\
            \bottomrule
            \end{tabular}}
            \caption{Priors used for the forwarded modeling. In the case of a Gaussian prior, we report the range at a $\pm 1\sigma$--level.}
            \label{tab:priors}
        \end{table}

        \section{Residuals}\label{app:resids}
        
\begin{figure}[h]
            \centering
            \includegraphics[width = 0.45\textwidth]{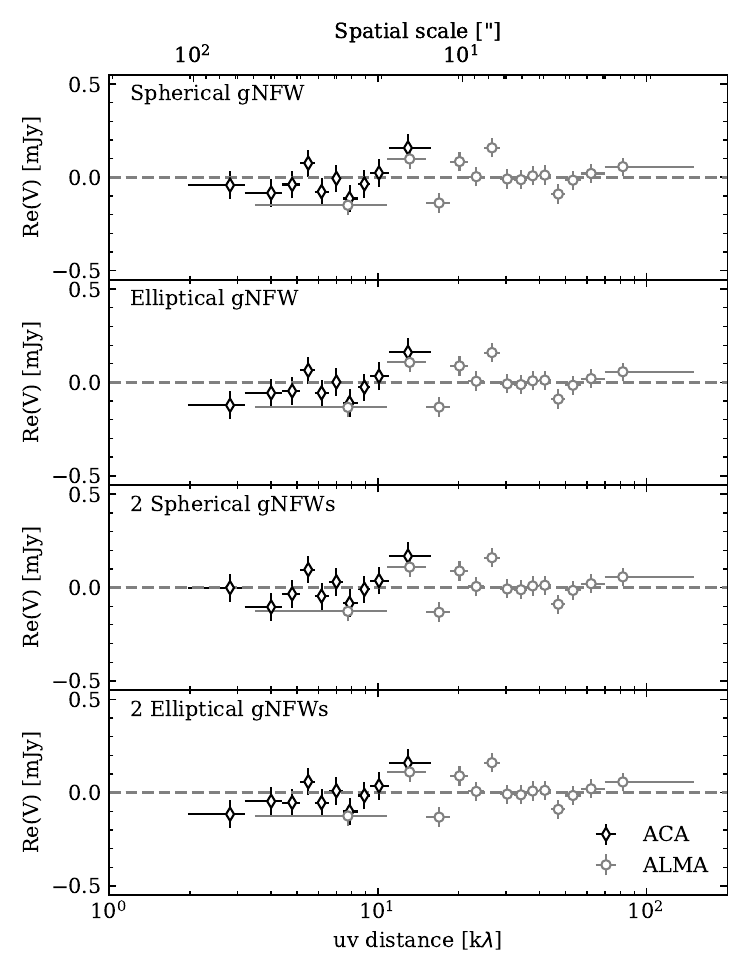}
            \includegraphics[width = 0.45\textwidth]{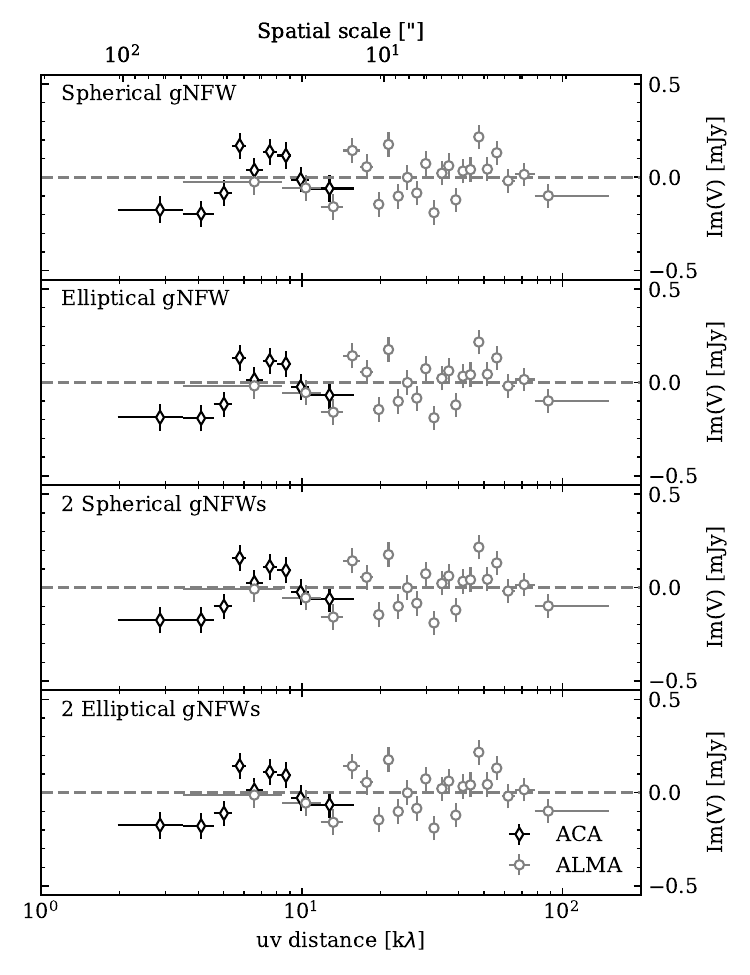}
            \caption{Residuals of the ALMA+ACA modeling in the \textit{uv}-plane for the single spherically symmetric gNFW ($\Delta |\ln{\mathcal{Z}}|=59.9$, top row), elliptical gNFW ($\Delta |\ln{\mathcal{Z}}|=62.0$, second row), two spherically symmetric components ($\Delta |\ln{\mathcal{Z}}|=62.3$, third row), and two elliptical gNFWs ($\Delta |\ln{\mathcal{Z}}|=68.6$, bottom row).}
            \label{fig:2comp_model-res}
        \end{figure}

        Figure~\ref{fig:2comp_model-res}%~\&~\ref{fig:2comp_model-res_2} 
        ~shows the residuals in the \textit{uv}-plane between the data, shown in Fig~\ref{fig:uv-radial}, and the likelihood weighted reconstruction of the spherical and elliptical gNFW models for when one (top two panels) and two components (bottom two panels) are used. We show the residuals from the ALMA+ACA modeling. If any remaining structure is present in XLSSC~122, such as local pressure perturbations or shock waves, they should be present in these residuals. Figure \ref{fig:2comp_model-res} shows the effect of different morphologies on the binned \textit{uv}-data, indicating that mapping these structures will be model-dependent and any small-scale residual will be indistinguishable from noise fluctuations because of the limited $S/N$ in the relevant spatial scales ($3-10~\rm k\lambda$).
        
\end{appendix}
\end{document}